\documentclass[structabstract]{aa}  
\usepackage{graphicx}
\usepackage{txfonts}
\usepackage{longtable}
\usepackage{lscape}

\usepackage{color}

\begin{document}
\title{Massive star formation in the GMC G345.5+1.0: Spatial  distribution of the dust emission}


   \author{Cristian L\'opez
          \inst{1}
          \and
          Leonardo Bronfman
          \inst{1}
          \and
          Lars-{\AA}ke Nyman
          \inst{2}
          \and
          Jorge May
          \inst{1}
          \and
          Guido Garay
          \inst{1}          
          }

   \institute{Departamento de Astronom\'{\i}a, Universidad de Chile,
              Casilla 36-D, Santiago, Chile,
              \email{clopez@das.uchile.cl}
           \and
             European Southern Observatory, Casilla 19001, Santiago, Chile.
   }
   \date{}

            
 \abstract 
{Massive condensations in giant molecular clouds (GMCs) are linked to
  the formation of high mass stars, which are the principal source of
  heavy elements and UV radiation, playing an important role in the
  evolution of galaxies.}
 { We attempt to make a complete census of massive-star formation within  all of
   GMC G345.5+1.0.  This cloud is located one degree above the
   galactic plane and at 1.8\,kpc from the Sun, thus there is little
   superposition of dust along the line-of-sight, minimizing confusion
   effects in identifying individual clumps.}
 {We observed the 1.2\,mm continuum emission across the whole GMC
   using  the Swedish-ESO Submillimetre Telescope\,(SEST) Imaging
       Bolometer Array\,(SIMBA) mounted on the SEST.  Observations
   have a spatial resolution of 0.2\,pc and cover
   1$_{^\centerdot}^\circ$8$\times$2$_{^\centerdot}^\circ$2 in the sky
   with a noise of 20\,mJy\,beam$^{-1}$.}
{We identify 201 clumps with diameters between 0.2 and 0.6\,pc, masses
  between 3.0 and 1.3$\times$10$^3$\,M$_\odot$, and densities between
  5$\times$10$^3$ and 4$\times$10$^5$\,cm$^{-3}$. The total mass of
  the clumps is 1.2$\times$10$^4$\,M$_\odot$, thus the efficiency 
  in forming
  these clumps, estimated as the ratio of the total clump mass to
  the total GMC mass, is $\sim$0.02.  The clump mass distribution for
  masses between 10 and 10$^3$\,$M_\odot$ is  well-fitted by a power
  law dN/dM$\propto$M$^{-\alpha}$, with a spectral mass index $\alpha$
  of 1.7$\pm$0.1.  Given their mass distribution, clumps do not appear
  to be the direct progenitors of single stars.  Comparing the 1.2\,mm
  continuum emission with infrared images taken by the Midcourse Space
  Experiment (MSX) and by the SPITZER satellite, we find that at least
  $\sim$20\% of the clumps are forming stars, and at most $\sim$80\%
  are starless.  Six  massive-star forming regions (MSFRs) embedded in
  clumps and associated with IRAS point sources have mean densities of
  $\sim$10$^5$\,cm$^{-3}$, luminosities  $>$10$^3$\,L$_\odot$, and
  spectral energy distributions that can be modeled with two dust
  components at different mean temperatures  of 28$\pm$5 and
  200$\pm$10\,K.}{}
   \keywords{ ISM: clouds -- ISM: stars: formation -- ISM: dust, extinction -- ISM: stars: circumstellar matter}

   \titlerunning{Massive star formation in the GMC G345.5+1.0}

   \maketitle
%

\section{Introduction}

\subsection{Giant molecular clouds}
\label{subsectionGiantMolecularClouds}

 High-mass stars are known to be born in massive and dense clumps 
embedded within giant molecular clouds (GMCs; Zinnecker \& Yorke 2007).
 These GMCs have typical radii of $\sim$60\,pc, masses of 10$^6$ M$_\odot$
and temperatures of $\sim$10\,K, and are found  to be concentrated 
toward the
Galactic plane (Grabelsky et al. 1988).  This gives rise to spatial
and kinematic blending  along the line-of-sight.  For example, the CO
emission, the most frequently used tracer of molecular gas for GMCs,
with a low critical density ($\sim$10$^2$\,cm$^{-3}$), is often
complex with multiple profile components.  To overcome these
observational difficulties and  perform a complete census of  massive-star
forming regions (MSFRs) in a whole GMC, ideally  requires one to
study GMCs  above and below the  Galactic plane using a high
density tracer.  Recent surveys of dust condensations within whole
GMCs have been made toward RCW 106 (Mookerjea et al. 2004), Cygnus X
 (Motte et al. 2007), and NGC 6334 (Mu\~noz et al. 2007) using
millimeter continuum emission as a high density tracer.

\subsection{Massive-star forming regions}

MSFRs are embedded in GMCs, generating a high Lyman continuum photon
flux ($\gtrsim$2$\times$10$^{45}$\,photons\,sec$^{-1}$; Panagia 1973) that ionizes the
surrounding gas and heats the surrounding dust. These MSFRs can thus be
identified by infrared emission from heated dust and/or by radio
emission from ionized gas.
They can also be identified  based on both their molecular lines from
high density gas and  their millimeter and sub-millimeter continuum
emission from dust.

Bronfman et al. (1996) showed that 60\% of the IRAS point sources in
the  Galactic plane with 
far infrared (FIR) colors typical of ultra
compact (UC) HII regions (Wood \& Churchwell 1989) are associated with
dense molecular structures seen in the CS(2-1) line
($\gtrsim$10$^4$-10$^5$\,cm$^{-3}$). Hereafter, these sources will be
 referred to as IRAS-CS sources.  
 From observations in 1.2\,mm continuum
emission of 146 IRAS-CS sources, Fa\'undez et al. (2004) showed that
MSFRs are associated with condensations of gas and dust.
Infrared studies, however,  cannot provide a
complete census of the birth sites of massive stars, since there are
massive condensations that are  undetected at infrared wavelengths.
For example, Garay et al. (2004) found four clumps with masses between
4$\times$10$^2$ and 2$\times$10$^3$\,M$_\odot$ and densities of
$\sim$2$\times$10$^5$\,cm$^{-3}$, without infrared emission, located
close to clumps associated with MSFRs.  They suggested that these cold
($\lesssim$17\,K),  dense, and massive clumps will eventually form high
mass stars.  
 Hill et al. (2005) found 113 cold clumps,  which 
have a mean mass of $\sim$800\,M$_\odot$, a mean radius of 
 $\sim$0.4\,pc, and a mean density of $\sim$10$^5$\,cm$^{-3}$.
Beltr\'an et al. (2006) 
found 95 cold clumps with 
a mean mass of 96\,M$_\odot$, a mean radius of 
 0.4\,pc, and a mean density of 9$\times$10$^4$ \,cm$^{-3}$.

\subsection{Clumps in GMCs}

Observations in molecular lines (e.g. Bains et al. 2006) and dust
continuum emission (e.g. Mu\~noz et al. 2007) on spatial scales
smaller than $\sim$1\,pc show that GMCs have a fragmented structure,
and these sub-structures have been  referred to as clumps (Williams et
al. 2000).

Clumps in GMCs have masses from 4 to 10$^4$\,M$_\odot$, diameters from
0.2 to  2\,pc, and densities from 10$^3$ to 10$^5$\,cm$^{-3}$.  The
clump mass distribution is consistent with a power law dN/dM$\propto$
M$^{-\alpha}$, where dN/dM is the number of objects by mass interval,
M is the  mass, and $\alpha$ is the mass spectral index.  The derived
mass spectral   indices range between 1.3 and 1.8 (e.g. Mookerjea et al. 2004; Mu\~noz et
al. 2007), values  similar to those found for 
the mass distribution of 
molecular clouds as a whole
($\alpha$=1.5-1.6; Sanders et al. 1985; Solomon et al. 1987; Williams
\& McKee 1997).  The similarity between the spectral mass   indices
suggests a common origin; however, different mechanisms have been
proposed to explain the formation of clumps and GMCs.  On the one
hand, large-scale gravitational instabilities, in the combined medium
of the collisionless stars and the collisional gas, drive spiral
density waves (e.g. Li et al. 2005), and are likely  to be the main mechanism behind
GMC formation.   On the other hand, 
since GMCs are considered turbulent and supersonic, 
it is expected that the clump formation  is, to first 
order, produced by ram pressures from supersonic flows, 
which can provide the seeds for a gravitational fragmentation
(Ballesteros-Paredes et al. 2006 and 2007; Bonnell et al. 2007; Klessen et al. 1998).

\subsection{This paper}

 To undertake a complete census of dense and massive clumps,
including those with and without infrared emission, we observed the
 whole of GMC G345.5+1.0 in 1.2\,mm continuum emission.  The determination
of physical properties of a sample of clumps belonging to a single GMC
is desirable since, in this way, the distance to  the GMC is a
factor that influences neither the mass distribution or the relationships
between physical properties.  However, these studies are difficult
because of the superposition of dust and molecular structures along the line-of-sight to GMCs in the
 Galactic plane.  We chose the GMC G345.5+1.0 as our target for two
reasons.  First, it is located $\sim$1$^\circ$  above the Galactic
plane, so our observations in the 1.2\,mm continuum emission 
and in the $^{13}$CO(3-2) line
are roughly
free of confusion with background or foreground structures  along the 
line-of-sight.  Second, it is at 1.8\,kpc from the Sun, near enough to
resolve clumps associated with  MSFRs, but far enough away to permit  a complete
coverage of the GMC.

The structure of this article is as follows:  Sections
\ref{sectionGMC_G345.5+1.0} and \ref{sectionObservations} present,
respectively, the main characteristics of G345.5+1.0 and the
observations;  Section 4 presents the results and a discussion, and  Sect.
\ref{sectionConclusions} gives a summary of our study.

\begin{figure}
\resizebox{\hsize}{!}{\includegraphics{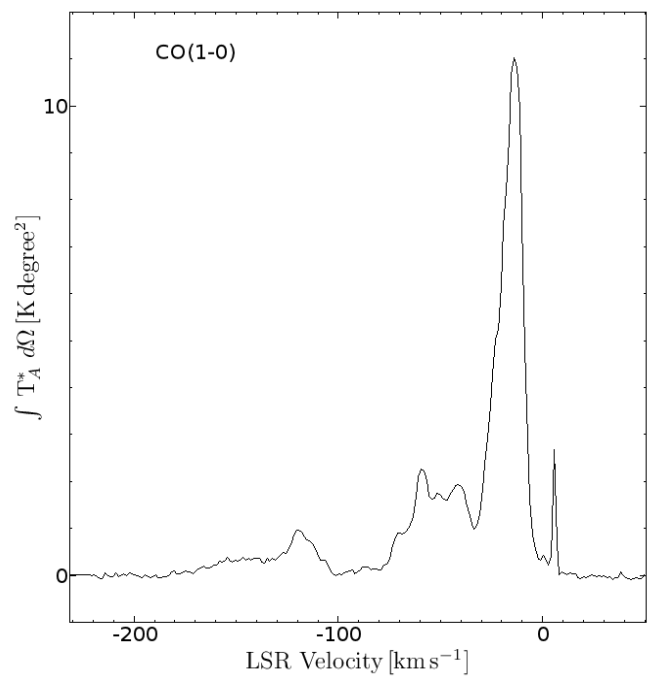}}
\caption{ Spectra of the $^{12}$CO(1-0) line emission integrated over
  the whole area of the GMC G345.5+1.0 (between
  344$_{^\centerdot}^\circ$5 and 346$_{^\centerdot}^\circ$5 in
   Galactic longitude and between 0$_{^\centerdot}^\circ$2 and
  2$_{^\centerdot}^\circ$0 in  Galactic latitude; Bronfman et
  al. 1989).  Emission from the GMC under study is between -33 and
  $-2$\,km\,s$^{-1}$ with a peak at $-13.6$\,km\,s$^{-1}$.}
\label{image12COEmission}
\end{figure}

\begin{figure*}
\centering
\resizebox{450pt}{!}{\includegraphics{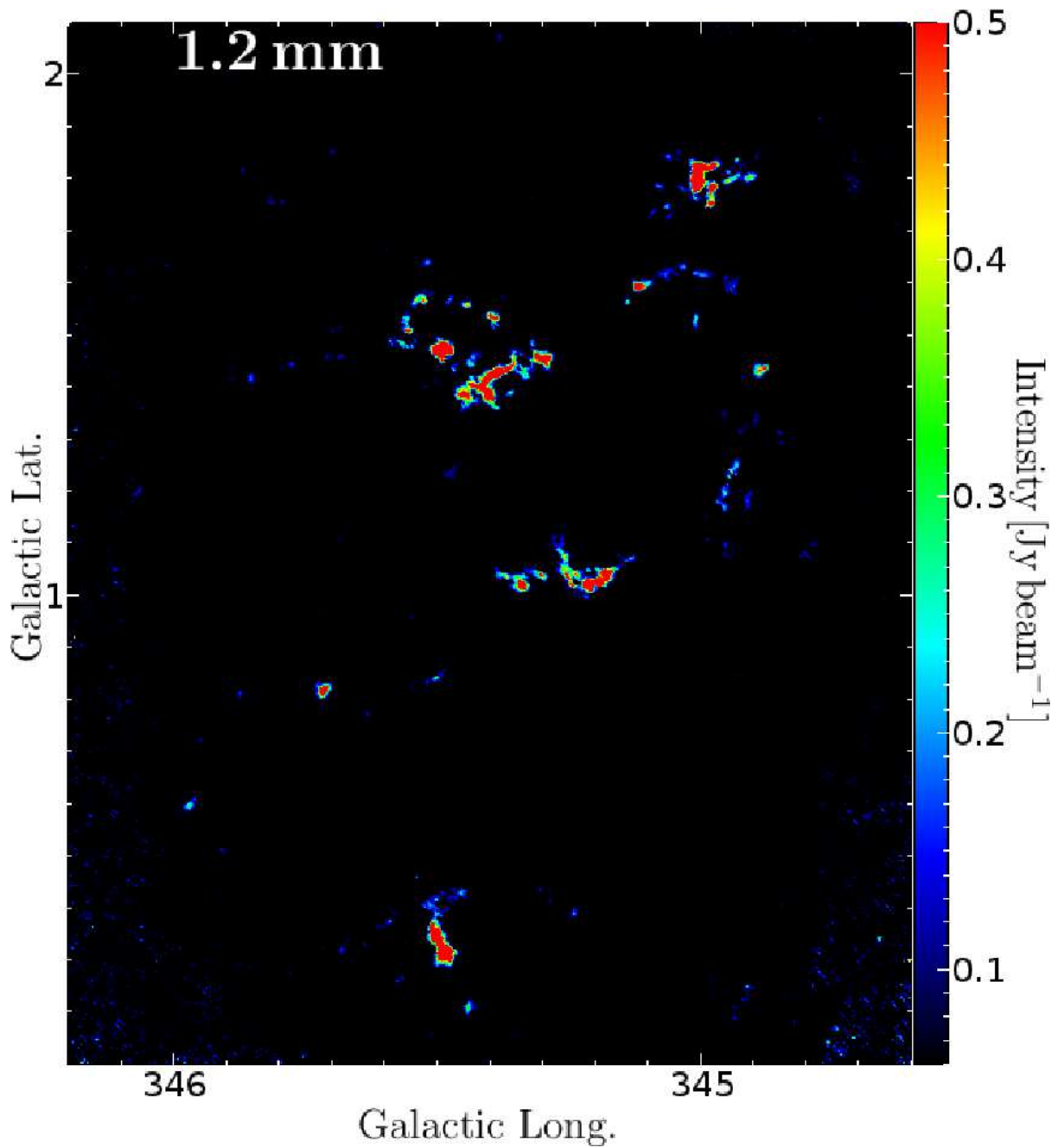}}
\caption{Map of the GMC G345.5+1.0 in 1.2\,mm continuum emission.
  Observations were made using SIMBA, with a spatial resolution of
  0.2\,pc.  They cover
  1$_{^\centerdot}^\circ$8$\times$2$_{^\centerdot}^\circ$2 in the sky,
  with an rms of 20\,mJy\,beam$^{-1}$.  }
\label{figureSIMBA}
\end{figure*}

\begin{center}
\begin{table}
\caption{Summary of main characteristics of  G345.5+1.0.}
\label{tableGMC}

\begin{tabular}{cc}
\hline\hline\\
Distance      &  1.8\,kpc \\
Total mass    & 6.5x10$^5$\,M$_\odot$\\
Radius        & 34\,pc \\
$^a$Density       & 70\,cm$^{-3}$\\
$^a$Column density& 10$^{22}$\,cm$^{-2}$\\
LSR velocity  & -13.6\,km\,s$^{-1}$ (between -33 and -2\,km\,s$^{-1}$)\\
\hline

\end{tabular}
\\
\begin{list}{}{}
 \item[$^a$]  The density and the column density are computed assuming a mean molecular
weight of $\mu$=2.29.
\end{list}{}{}

\end{table}
\end{center}

\section{GMC G345.5+1.0}
\label{sectionGMC_G345.5+1.0}

The GMC G345.5+1.0 was first observed  as part of the Columbia University -
Universidad de Chile $^{12}$CO(1-0) Survey of the Southern Galaxy
(Bronfman et al. 1989).  It is located approximately between
344$_{^\centerdot}^\circ$5 and 346$_{^\centerdot}^\circ$5 in Galactic
longitude, and between 0$_{^\centerdot}^\circ$2 and
2$_{^\centerdot}^\circ$0 in  Galactic latitude.
 Figure \ref{image12COEmission} shows the spectrum of the $^{12}$CO(1-0)
line emission integrated over the area of the GMC.  The emission of
the GMC is between $-33$ and $-2$\,km\,s$^{-1}$ (LSR velocities) with a peak
at $-13.6$\,km\,s$^{-1}$.

We estimate the kinematic distance using  the rotation
curve determined by Alvarez et al. (1990), with a  Galactocentric solar
distance of 8.5\,kpc and a solar LSR velocity of 220\,km\,s$^{-1}$.
Considering a  Galactic longitude of 345$_{^\centerdot}^\circ$5 and LSR
velocity of $-13.6$\,km\,s$^{-1}$, the GMC is within the solar circle with two
possible kinematic distances: $\sim$1.8 and 15\,kpc.  Thus, the GMC is
$\sim$31 or 262\,pc above the  Galactic plane, a factor 0.5 or 4.4 of
the HWHM of the molecular  Galactic disk ($\sim$60\,pc; Bronfman et
al. 2000), respectively. Therefore, 1.8\,kpc is the most probable
kinematic distance to the GMC.

Physical properties of the GMC are estimated using the $^{12}$CO(1-0)
line observations.  The $^{12}$CO(1-0) line emission integrated over
the full spatial and spectral extension of the GMC is 192
 K\,km\,s$^{-1}$\,deg$^2$.  Using a ratio of H$_2$ column density to
integrated $^{12}$CO(1-0) line emission $N_{H_2}/W_{co}$ equal to
1.56x10$^{20}$\,cm$^{-2}$\,(K\,km\,s$^{-1}$)$^{-1}$ (Hunter et
al. 1997), the total mass of the GMC is 6.3x10$^5$\,M$_\odot$,
corrected by a factor of 1.3 to account for 25\% of helium.
 Since its angular size is
2$_{^\centerdot}^\circ$0x1$_{^\centerdot}^\circ$8, the GMC has a mean
radius of $\sim$34\,pc, mean column density of
$\sim$10$^{22}$\,cm$^{-2}$ and mean density of $\sim$70\,cm$^{-3}$,
 where the depth of the cloud has been assumed to be same as its
  radius. Table \ref{tableGMC} summarises the main characteristics of
this GMC.  The derived physical properties confirm that the most
probable distance to the GMC is $\sim$1.8\,kpc.  If at the far
kinematic distance (15\,kpc), the GMC would have a mean radius of
280\,pc and a total mass of 4.4x10$^7$\,M$_\odot$, values much higher
than the  typical value of 60\,pc in radius and 10$^6$\,M$_\odot$ in mass
(Dame et al. 1986); Williams \& McKee (1997) found that the molecular
cloud mass distribution within the solar circle has an upper mass
limit of 6$\times$10$^6$\,M$_\odot$.

\section{Observations}
\label{sectionObservations}

We observed the whole GMC G345.5+1.0 in continuum emission at 1.2\,mm
using SIMBA mounted on the
SEST.   The SEST is a 15-m diameter
radio telescope,  which operated between 70 and  365\,GHz, and SIMBA is a
 37-channel hexagonal bolometer, operating at 250\,GHz (1.2\,mm), with
a passband equivalent width of 90\,GHz (FWHM).  The configuration
SEST-SIMBA had a beamsize of 24$''$, which corresponds to a spatial
resolution of $\sim$0.2\,pc at the GMC distance (1.8\,kpc).

 Observations were made in October 2002  and July 2003, using the
 fast mapping  mode, and consist of 185 images of 15$'$ (azimuth)
 $\times$10$'$ (elevation) in size.  The scans were made in
 azimuth at a rate of 80$''$\,s$^{-1}$, and they were separated in
 elevation by 8$''$; the total integration time per map was about 25
 minutes.   Measurements of the atmospheric opacity were made
   through skydips about every  three  hours, and values at the zenith ranged
   between 0.09 and 0.31.  Data were reduced  using the MOPSI software
 (developed by Robert Zylka, IRAM, Grenoble, France), and calibrated
  (in terms of flux density) with observations toward Uranus, one made in
 October 2002, and  an additional eight made in July 2003.  Bolometer channels were
   corrected for the correlated noise by inspecting the surrounding
   channels.  The noise correlation between 1 and 900\,arcsec, and
   between 100 and 900\,arcsec 
    inferred an uncertainty/error smaller
   than 20\% in the
   flux densities  of detected sources.  The calibration factor has a
   value of 0.086\,Jy\,counts$^{-1}$ for October 2002, and a mean value
   of $\sim$0.069$\pm$0.005\,Jy\,counts$^{-1}$ for July 2003.  
    These
   variations agree with the uncertainty
   estimated by Fa\'undez et al. (2004)
   of 20\% in the  flux
   density measurements using SIMBA data.

For the 185 images, we achieved an rms of between 0.023 and
0.080\,Jy\,beam$^{-1}$, with a median of 0.036\,Jy\,beam$^{-1}$, for a
beam calibration area of $\sim$653\,arcsec$^2$ (``SIMBA Data Reduction
Handbook'').  The individual maps were combined in a mosaic of
1$_{^\centerdot}^\circ$8$\times$2$_{^\centerdot}^\circ$2 in size,
centered at 345$_{^\centerdot}^\circ$40 in  Galactic longitude and
+1$_{^\centerdot}^\circ$10 in  Galactic latitude, and with a final rms
of $\sim$20\,mJy\,beam$^{-1}$.

\begin{figure}
\resizebox{240pt}{!}{\includegraphics{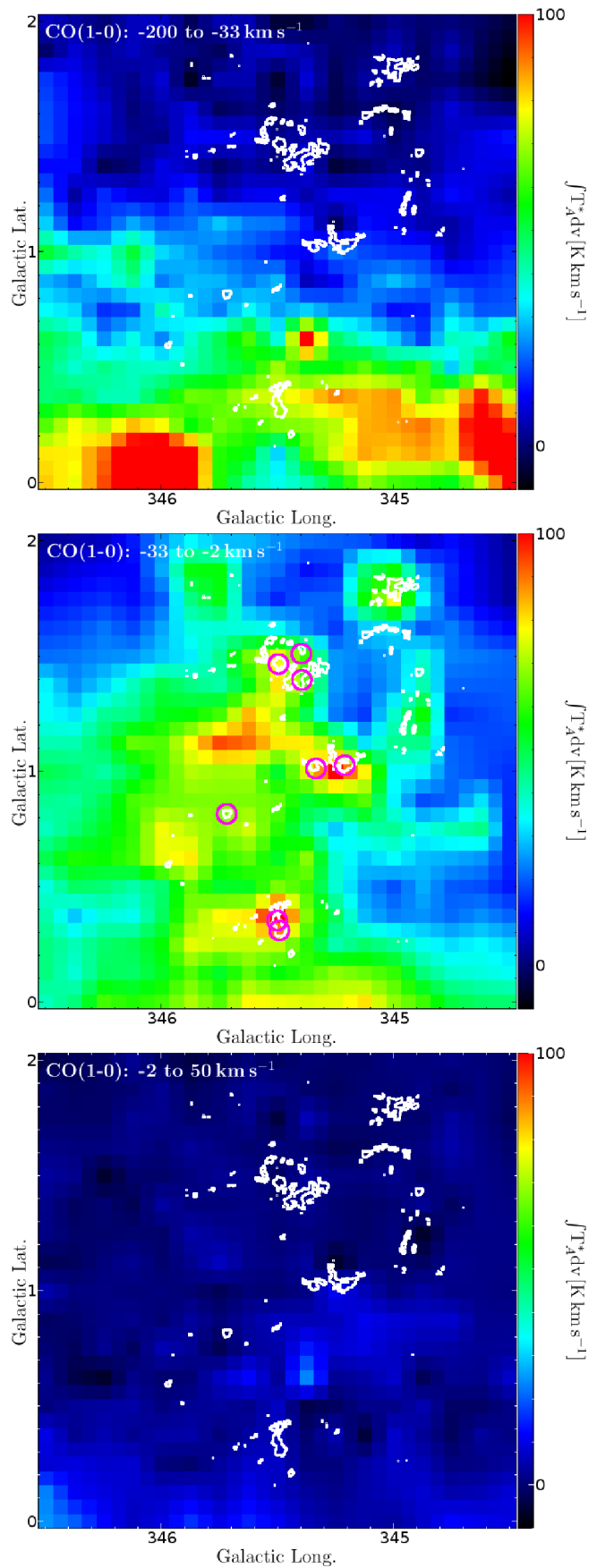}}
\caption{Integrated $^{12}$CO(1-0) emission toward
  GMC G345.5+1.0 in different LSR velocity ranges (Bronfman et
  al. 1989).  Top: from $-200$ to $-33$\,km\,s$^{-1}$.  Middle: from $-33$ to
  $-2$\,km\,s$^{-1}$. Bottom: from $-2$ to 50\,km\,s$^{-1}$.  Magenta circles mark spatial
  and spectral positions of detections in the CS(2-1) line toward
  MSFRs (Table \ref{tableCS}).  Contours represent 1.2\,mm continuum
  emission at 5 times rms, $\sim$0.1\,Jy\,beam$^{-1}$.}
\label{figureCOChannelMaps}
\end{figure}

\begin{center}
\begin{table}
\caption{List of IRAS point sources 
 along the line-of-sight to the GMC G345.5+1.0
observed 
in the CS(2-1) line by Bronfman et al. (1996).}
\label{tableCS}
\begin{tabular}{ccccc}
\hline
\hline\\
IRAS name &\multicolumn{2}{c}{Galactic coord.} & LSR velocity$^1$ & FWHM$^1$ \\
          &longitude      & latitude        &  [km\,s$^{-1}$]      &[km\,s$^{-1}$]\\
\hline\\
16533-4022& 344.845 & 1.646 &\multicolumn{2}{c}{undetected}\\
16571-4029& 345.208 & 1.028 & -15.6& 4.9 \\
16577-4028& 345.286 & 0.933 &\multicolumn{2}{c}{undetected}\\
16575-4023& 345.332 & 1.014 & -14.5& 2.8\\ 
16561-4006& 345.393 & 1.399 & -11.9& 3.5\\
16557-4002& 345.395 & 1.512 & -12.4& 3.1\\ 
17009-4042& 345.490 & 0.311 & -16.7& 6.0\\ 
16562-3959& 345.494 & 1.468 & -11.6& 5.5\\ 
17008-4040& 345.499 & 0.354 & -16.4& 4.8\\ 
16596-4012& 345.717 & 0.817 & -11.5& 4.9\\ 
\hline
\end{tabular}
\begin{list}{}{}
\item[$^1$] Values from Bronfman et al. (1996).
\end{list}{}{}
\end{table}
\end{center}

\begin{figure}
  \resizebox{\hsize}{!}{\includegraphics{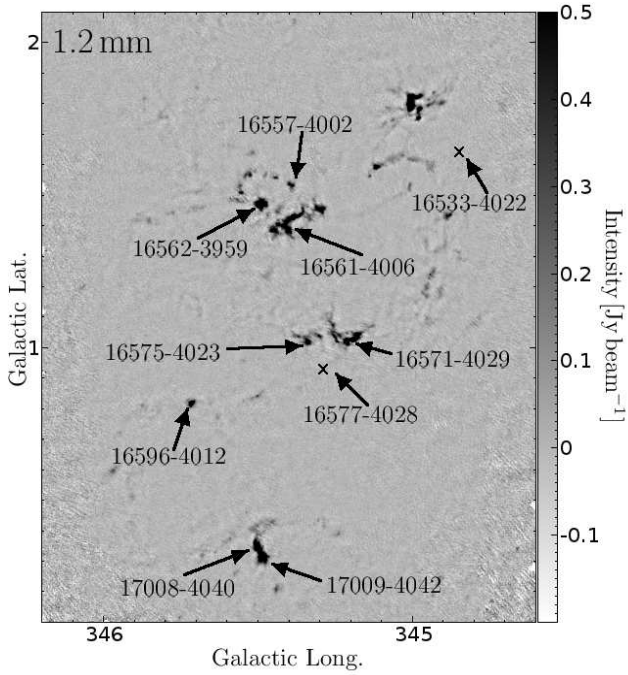}}
  \caption{ IRAS point sources  along the line-of-sight of the GMC
    G345.5+1.0 observed in the CS(2-1) line (Bronfman et al. 1996).  Gray
    scale represents 1.2\,mm continuum emission.  Arrows mark
    CS(2-1) line observations, and crosses indicate observations
    without detection (see Table \ref{tableCS}).}
\label{imageSIMBA-CS}
\end{figure}

\section{Results and discussion}
\label{sectionResults}

\subsection{ The 1.2\,mm continuum emission}

\begin{figure}
  \resizebox{250pt}{!}{\includegraphics{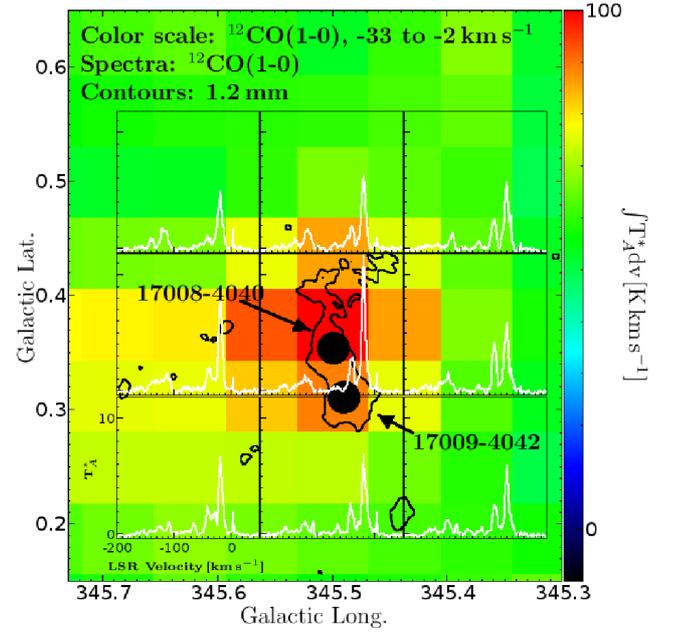} }\\
  \resizebox{250pt}{!}{\includegraphics{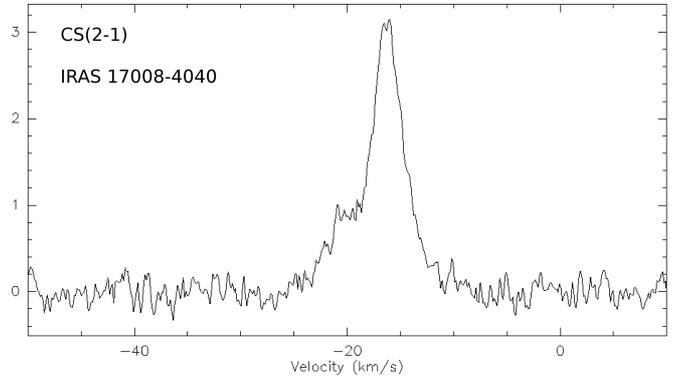} }\\
  \resizebox{250pt}{!}{\includegraphics{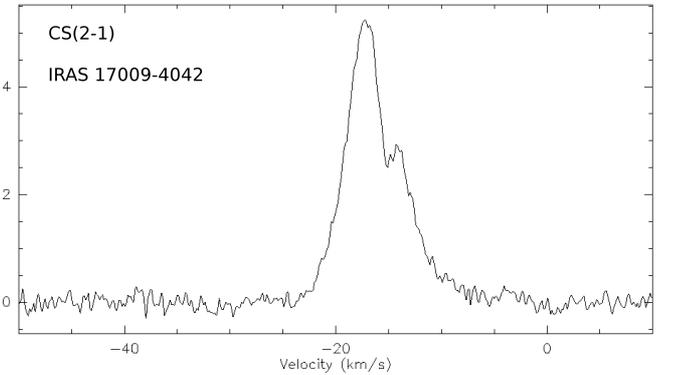} }
  \caption{Line profiles toward the IRAS point sources 17008-4040 and
    17009-4042: top image, $^{12}$CO(1-0) line profiles (Bronfman et
    al. 1989), over a map of their integrated emissions (color scale)
    and with contours of 1.2\,mm continuum emission; middle and bottom
    images, CS(2-1) line profiles (their observing positions are
    indicated as black dots in the top image; Bronfman et al. 1996).}
\label{imageSIMBA-CS-CO}
\end{figure}

 Figure \ref{figureSIMBA} presents the 1.2\,mm emission image of the
 whole GMC G345.5+1.0.  The total flux density of the GMC G345.5+1.0
 is $\sim$365\,Jy, which is estimated by integrating the intensity
 over the  whole area of the cloud.

As can be seen in the integrated spectrum of the $^{12}$CO$(1-0)$ line
(Fig. \ref{image12COEmission}), across GMC G345.5+1.0 (LSR velocity
between $-33$ and $-2$\,km\,s$^{-1}$), there are  additional molecular gas
components along the  line-of-sight, particularly in the ranges $-170$
to $-100$\,km\,s$^{-1}$, $-75$ to $-33$\,km\,s$^{-1}$, and 0 to
5\,km\,s$^{-1}$.  Hence, the question arises as to whether the 1.2\,mm
continuum emission only traces dust condensations within GMC
G345.5+1.0?  To examine the association of the GMC with 1.2\,mm
continuum emission, Fig. \ref{figureCOChannelMaps} shows images of the
velocity-integrated $^{12}$CO$(1-0)$ emission in three velocity ranges
($-200$ to $-33$ km\,s$^{-1}$, $-33$ to $-2$  km\,s$^{-1}$, and $-2$ to
$50$ km\,s$^{-1}$) superimposed with contours of the 1.2\,mm continuum
emission.  Figure \ref{figureCOChannelMaps} shows that most of the
emission detected in 1.2\,mm is associated with the GMC.  About 1\% of
the total observed area might  also be associated with gas at
velocities $<$$-33$\,km\,s$^{-1}$ (Fig. \ref{figureCOChannelMaps},
top), localized mainly in the region of the IRAS point sources
17008-4040 (G345.499+0.354) and 17009-4042 (G345.490+0.311).

From the survey of Bronfman et al. (1996), we find that there are  eight
IRAS-CS sources within the region and two IRAS point sources  that are not
detected in the CS(2-1) line (see Table \ref{tableCS}).  As is
shown in  both the $^{12}$CO(1-0) maps (Fig. \ref{figureCOChannelMaps})  and
the 1.2\,mm continuum emission map (Fig. \ref{imageSIMBA-CS}), these
MSFRs, or IRAS-CS sources, are associated with the GMC and have a
counterpart in 1.2\,mm. They correspond to the most dense and massive
dust condensations (see Table \ref{tableClumps}).  The two IRAS point
sources not detected in the CS(2-1) line were also not detected in the
continuum (see Fig. \ref{imageSIMBA-CS}).  The  eight IRAS-CS sources
include the IRAS point sources 17008-4040 and 17009-4042.  Line
profiles toward these two objects in the $^{12}$CO(1-0) and CS(2-1)
lines are shown in Fig. \ref{imageSIMBA-CS-CO}.  Gas components with
velocities $<$$-33$\,km\,s$^{-1}$ observed in the $^{12}$CO(1-0) line are not
observed in the CS(2-1) line, suggesting that they correspond to
regions of low density gas.  Since 1.2\,mm continuum emission traces
high densities (e.g. Fa\'undez et al. 2004), it should not be
detected in these clouds. In summary,
from observations in the $^{12}$CO(1-0) and CS(2-1) lines, we conclude that the
1.2\,mm continuum emission is associated only with the GMC.\\


\subsection{Identification of clumps}

 The structure of the GMC observed in 1.2\,mm continuum emission
(Fig. \ref{figureSIMBA}) is fragmented, and it is possible to
distinguish several clumps.

 To identify clumps we utilize  CLUMPFIND\footnote{http://www.ifa.hawaii.edu/users/jpw/clumpfind.shtml} (Williams et al. 1994), which
 creates contours over data, searches for peaks of emission to locate
 clumps, and follows them down to the lower intensity contour.

  CLUMPFIND finds 201 clumps in the 1.2\,mm continuum emission map
  of the GMC, containing $\sim$100\% of the  total emission above 3$\sigma$.
  We used a lower intensity contour of three rms, $\sim$0.06 Jy\,beam$^{-1}$,
  and a contouring interval equal to twice the rms noise, $\sim$0.04 Jy\,beam$^{-1}$.
   To
  delete fictitious structures, we imposed two conditions on   
  the CLUMPFIND output, that
  the angular size of the emission and the emission peak
  of clumps had to be greater than the beam size,
  $\sim$24$''$$\times$24$''$, and five times rms,
  $\sim$0.1\,Jy\,beam$^{-1}$, respectively.  
  The angular size  is defined to be  the angular area  inside the
  lowest intensity contour (three rms).
  The 201 identified clumps
  have  areas  between $\sim$0.18 and 7.3\,arcmin$^2$,
  emission peaks between  0.1 and 9\,Jy\,beam$^{-1}$, and flux densities between 0.089 and 40\,Jy.

 Table \ref{tableClumps},  available online in electronic form 
at  the publishers\footnote{http://www.edpsciences.org} (EDP Sciences), shows the
  characteristics of each clump calculated in this section,
   Sect. \ref{subsectionPhysicalPropertiesOfClumps}
  and
   Sect. \ref{sectionAssociationWithInfraredEmission}.  Column 1 gives
  clump names; columns 2 and 3,  Galactic coordinates of peaks in 1.2
  mm continuum emission; column 4, 1.2 mm flux densities; column 5,
  diameters (deconvolved FWHM sizes); column 6, masses;
  column 7, densities; column 8, column densities; and column 9, if
  clumps have ("Y") or do not have ("N") an infrared counterpart from
  MSX and SPITZER observations.

\subsection{Physical properties of clumps}
\label{subsectionPhysicalPropertiesOfClumps}

First, we estimate the minimum gas column density that can be detected given
the rms noise of our observations.  Assuming that 1.2\,mm continuum
emission is optically thin and produced by dust, the column density
$N$ is (Hildebrand 1983)
\begin{equation}
\label{equationN}
\qquad\qquad N= \frac{S_{1.2\,mm}\,R_{gd}}
{\Omega\ \mu\ m_H\ k_{1.2\,mm}\ B_{1.2\,mm}(T_{dust})},
\end{equation}
where $\Omega$ is the beam solid angle, $S_{1.2\,mm}$ is the
flux density at 1.2\,mm, $\mu$ is the mean mass per particle,
 equal to $\sim$2.29 for  an H$_2$ cloud with a 25\%  contribution of
helium (Evans 1999), $m_H$ is the hydrogen atom mass, $k_{1.2\,mm}$ is
the dust absorption coefficient at 1.2\,mm,  equal to $\sim$1\,cm$^2$\,g$^{-1}$
for protostellar cores (Ossenkopf \& Henning 1994),
$B_{1.2\,mm}(T_{dust})$ is the Planck function at  both 1.2\,mm and  a dust
temperature $T_{dust}$, equal to  $\sim$30\,K for regions of  massive-star
formation (Fa\'undez et al. 2004), and $R_{gd}$ is the ratio of gas to
dust mass, $\sim$100 (Hildebrand 1983).   For a solid angle
limit of 24$''$$\times$24$''$ and an intensity limit of five rms,
$\sim$0.1\,Jy\,beam$^{-1}$, the minimum flux density is $\sim$88\,mJy
and the minimum column density that can be detected is
$\sim$4x10$^{21}$\,cm$^{-2}$, which corresponds to a visual extinction
of 4\,mag, assuming  that $N_{H_2}/A_V$$\sim$$10^{21}cm^{-2}\,mag^{-1}$
(Bohlin et al. 1978).

Masses of clumps, M$_c$, are estimated as
$$M_c = \int \mu\,m_H N dA =
\frac{S_{1.2\,mm}\ R_{gd}\ d^2}{B_{1.2\,mm}(T_{dust})\ k_{1.2\,mm}},$$
where dA is the differential element of area (dA= d$^2$\,d$\Omega$)
and $d$ is the distance to the GMC ($\sim$1.8\,kpc).  Since we  insist
that  identified  clumps
have intensities and dimensions greater than
0.1\,Jy\,beam$^{-1}$ and  24$''$$\times$24$''$ respectively, the lower
limit to their masses is $\sim$2.9\,M$_\odot$.  The derived masses of
clumps range from 3.0 to 1.3$\times$10$^3$\,M$_\odot$, with a total
mass of 1.2x10$^4$\,M$_\odot$ (see Tables \ref{tableSummaryClumps} and
\ref{tableClumps}).  The efficiency in forming these clumps, estimated as
the ratio of the total clump mass to the total GMC mass, is thus
$\sim$0.02.

\begin{figure}
\resizebox{\hsize}{!}{\includegraphics{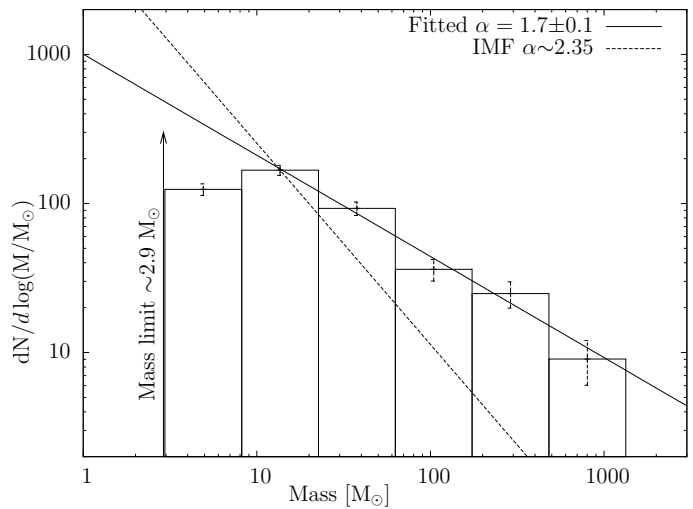}}
\caption{ Mass distribution of identified clumps in G345.5+1.0, plotted
  as dN/d$\log$(M/M$_\odot$) versus mass, where
  dN/d$\log$(M/M$_\odot$) is approximated by the number of clumps
  $\Delta$N within a logarithmic mass interval
  $\Delta\log($M/M$_\odot)$.  Here, $\Delta\log($M/M$_\odot)$ is
  constant,$\sim$0.44.  
  Error bars are estimated by
   $\sqrt{\Delta N/\Delta\log(M/M_\odot)}$.
  The arrow shows the clump mass limit, $\sim$2.9\,M$_\odot$.
   The continuous line represents  the mass distribution fit with
  $dN/d\log(M/M_\odot)$$\propto$M$^{1-\alpha}$, where the spectral mass index
  $\alpha$ is 1.7$\pm$0.1 for masses between $\sim$10 and
  1.3x10$^3$\,M$_\odot$.   The dashed line displays the spectral
  mass  index for the stellar initial mass function (IMF) of the solar
  neighborhood for stellar masses greater than 0.5\,M$_\odot$
   (e.g. Kroupa 2002);  the line is
  forced to pass through the peak of the clump mass distribution.}
\label{figureCMD}
\end{figure}

\begin{figure}
\resizebox{\hsize}{!}{\includegraphics{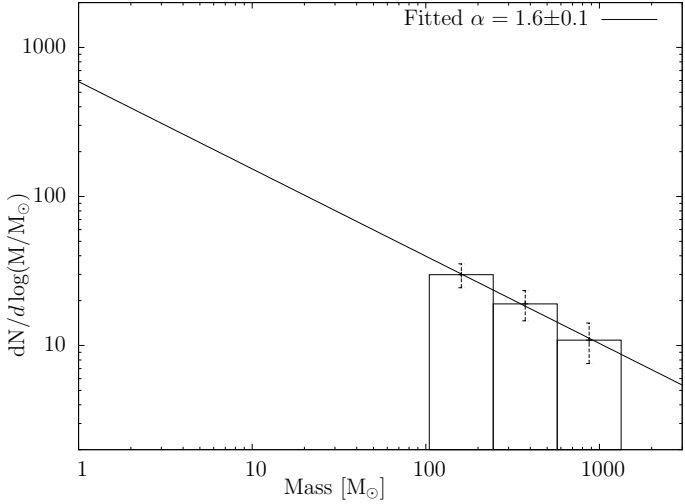}}
\caption{ Mass distribution of identified clumps in G345.5+1.0 with
  masses  higher than 100\,M$_\odot$, plotted as
  dN/d$\log$(M/M$_\odot$) versus mass, where dN/d$\log$(M/M$_\odot$)
  is approximated by the number of clumps $\Delta$N within a
  logarithmic mass interval $\Delta\log($M/M$_\odot)$.  Here,
  $\Delta\log($M/M$_\odot)$ is constant,$\sim$0.37.  Error bars are
  estimated by $\sqrt{\Delta N/\Delta\log(M/M_\odot)}$.   The continuous
  line represents  the mass distribution fit with
  $dN/d\log(M/M_\odot)$$\propto$M$^{1-\alpha}$, where the spectral
  mass index $\alpha$ is 1.6$\pm$0.1 for masses between $\sim$100 and
  1.3x10$^3$\,M$_\odot$.}
\label{figureCMDMassive}
\end{figure}

  The clump mass distribution (CMD) is shown in Fig. \ref{figureCMD},
  plotted as $dN/d\log(M/M_\odot)$ versus mass, where
  $dN/d\log(M/M_\odot)$ is approximated by the number of clumps,
  $\Delta$N, within a logarithmic mass interval
  $\Delta\log(M/M_\odot)$.  In  this figure, $\Delta\log(M/M_\odot)$ is
  constant,  at a value of $\sim$0.44.  The CMD is  well-fitted by a power law
  $dN/d\,log(M)\propto$M$^{-\alpha+1}$, which can be expressed as
  $dN/dM\propto$M$^{-\alpha}$, with the spectral mass index, $\alpha$,
  equal to 1.7$\pm$0.1 for masses between $\sim$10 and
  1.3x10$^3$\,M$_\odot$.  The correlation coefficient of the fit is
  0.993.  Between 10\,M$_\odot$ and 1.3x10$^3$\,M$_\odot$,  bin size
  variations of $\Delta\log(M/M_\odot)$ between 0.18 and 1.1 result in
  values of $\alpha$ consistent with 1.7$\pm$0.1.  Since $\alpha$ is
  1.7, the population is dominated by clumps with low masses, but the
  total mass is dominated by the most massive clumps; for example,
  50\% of the population is between 10 and 27\,M$_\odot$, but contains
  only 10\% of the total mass.  The turnover below $\sim$10\,M$_\odot$
  is produced by an incompleteness of the clump catalog  caused by the
  combination of the spatial resolution and flux density limit of the
  survey.  However, observations of higher spatial resolution
  ($\lesssim$0.01\,pc) could result in a spectral mass index of
  $\sim$2.35, resolving core structures (e.g. Motte et al. 1998).  
 Beltr\'an et al. (2006) studied a sample of IRAS sources
    associated with MSFRs in 1.2\,mm continuum and found a spectral
    mass index of 1.5 for clumps with masses between $\sim$10\,M$_\odot$ and
    10$^2$\,M$_\odot$ and 2.1 for clumps with masses between
    $\sim$10$^2$ and 10$^4$\,M$_\odot$.  For clumps identified here
    with masses higher than 100\,M$_\odot$, we find a spectral mass
    index of 1.6$\pm$0.1, as is shown in Fig.  \ref{figureCMDMassive},
    in agreement with the previous fit considering clumps with masses
    between 10 and 1.3$\times$10$^3$\,M$_\odot$.  
    It is necessary to observe more whole GMCs to confirm these results.

 Clump diameters, $D_c$, are estimated from  
the deconvolved FWHM size of their emissions.
We used the FWHM size $\theta_{FWHM}$ estimated by CLUMPFIND algorithm, thus 
$$D_c =\sqrt{ \theta_{FWHM}^2 - \theta_{beam}^2},$$ where
$\theta_{beam}$ is the beam-size.  Considering a distance of 1.8\,kpc
to the GMC and clumps that have a reliable $D_c$, i.e.
$D_c$$\ge$$\theta_{beam}$, clumps have diameters between 0.2 and
0.6\,pc.

From the masses and sizes, and assuming a spherical and homogeneous
density distribution, we estimate mean clump densities, using the
expression
$$\rho=\mu\ m_H\ n,$$ where $\rho$ is the mass  density and $n$ is the
 particle density.  Densities of clumps are between
5$\times$10$^3$ and 4$\times$10$^5$\,cm$^{-3}$.  Mean column densities
of clumps, $N_c$, are estimated as
$$N_c \sim \frac{M_c}{\mu\,m_H\,\pi(D_c/2)^2},$$ and range between
  4$\times$10$^{21}$ and 4$\times$10$^{23}$\,cm$^{-2}$.  Tables
\ref{tableSummaryClumps} and \ref{tableClumps} show physical
properties for each clump and a summary of them, respectively.

\begin{figure*}
\begin{center}
\resizebox{\hsize}{!}{\includegraphics{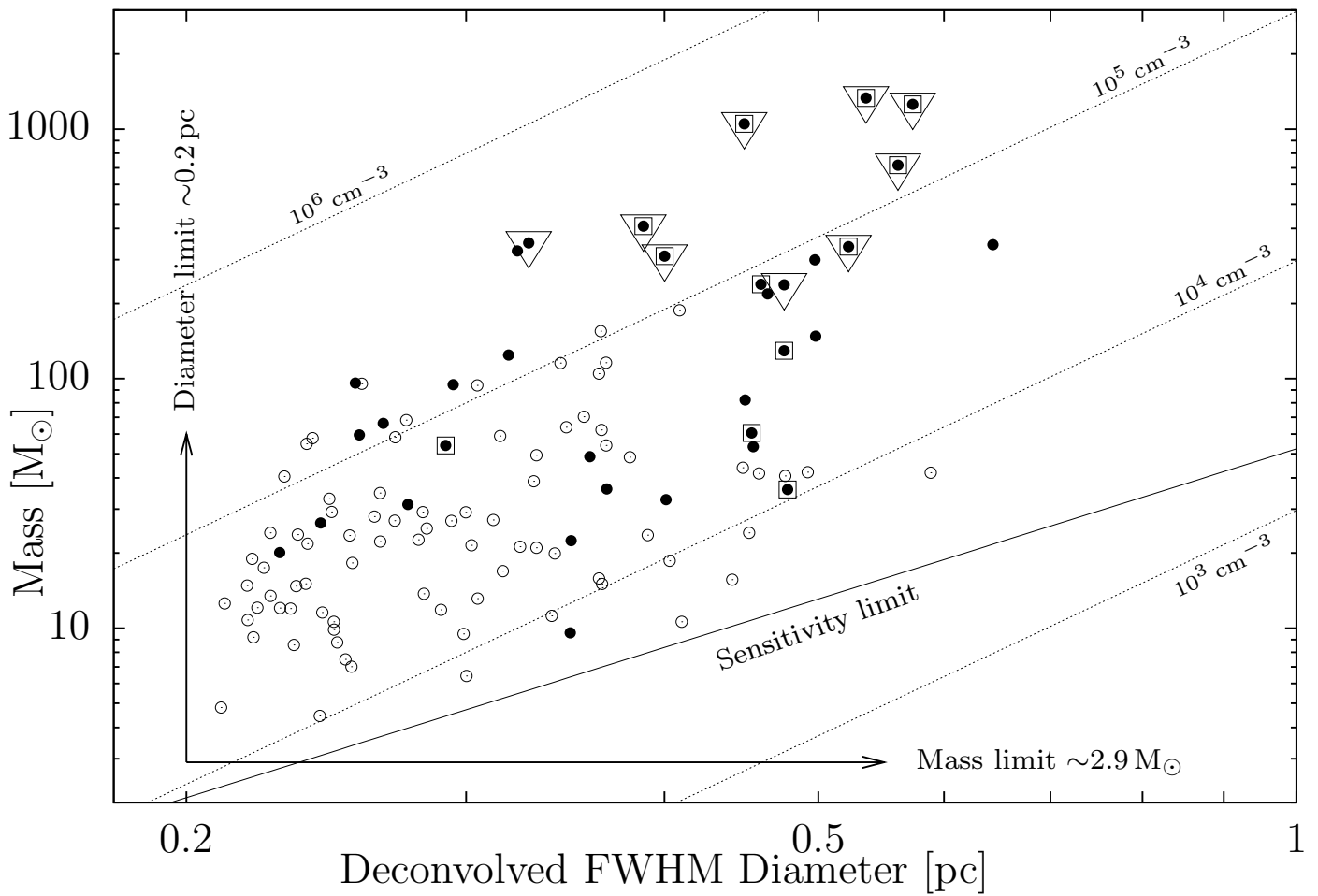}}\\

\end{center}
\caption{ Mass versus diameter for the clumps detected toward the
  GMC G345.5+1.0 in 1.2\,mm continuum emission with reliable
  diameters.  Filled circles indicate clumps detected in infrared MSX
  and SPITZER bands.  Open circles indicate clumps that do not have an
  infrared counterpart.  Triangles indicate clumps associated with
  MSFR-IRAS sources, which have luminosities $>$10$^3$\,L$_\odot$.
  Boxes indicate clumps associated with MSX sources that satisfy MYSO
  candidate criterion (Lumsden et al. 2002).  Arrows mark detection
  limits for masses ($\sim$2.9\,M$_\odot$) and diameters
  ($\sim$0.2\,pc).  
    The
  continuous line indicates the detectable mass as a function of 
  diameter (sensitivity limit),
  considering an intensity limit of five rms
   ($\sim$0.1\,Jy\,beam$^{-1}$). 
  Dotted lines indicate mean densities
  at 10$^3$, 10$^4$, 10$^5$ and  10$^6$\,cm$^{-3}$.
 The densities are computed  assuming 
a mean molecular weight of $\mu$=2.29.}
\label{figureSize-Mass}
\end{figure*}

Figure \ref{figureSize-Mass} shows a plot of mass versus diameter for
the clumps detected toward GMC G345.5+1.0 with reliable diameters.
The dotted lines indicate constant densities at 10$^3$, 10$^4$,
 10$^5$, and 10$^6$\,cm$^{-3}$.  The majority of clumps have
densities between 10$^4$ and 10$^5$\,cm$^{-3}$.

The physical properties of detected clumps are similar to those found
in other GMCs (e.g. Mookerjea et al. 2004).

\begin{center}
\begin{table}
\caption{Summary of the physical properties of the identified clumps.}
\label{tableSummaryClumps}
\centering
\begin{tabular}{ccccc }
\hline\hline\\

             &S$_{1.2\,mm}$ & Diameters & Masses       & $^a$Densities  \\
             &Jy               &pc         &M$_\odot$     & cm$^{-3}$       \\
             \hline\\

Range        &0.089-40            &0.2-0.6   &3.0-1.3$\times$10$^3$ &5$\times$10$^3$-4$\times$10$^5$ \\
Total        &3.7$\times$10$^2$ &          &1.2$\times$10$^4$    \\

\hline

\end{tabular}
\begin{list}{}{}
\item[$^a$]  The densities are computed  assuming 
a mean molecular weight of $\mu$=2.29.
\end{list}{}{}
\end{table}
\end{center}

\subsection{Association with infrared emission (IRAS-MSX-SPITZER)}
\label{sectionAssociationWithInfraredEmission}

\begin{figure}[!h]
  \resizebox{\hsize}{!}{\includegraphics{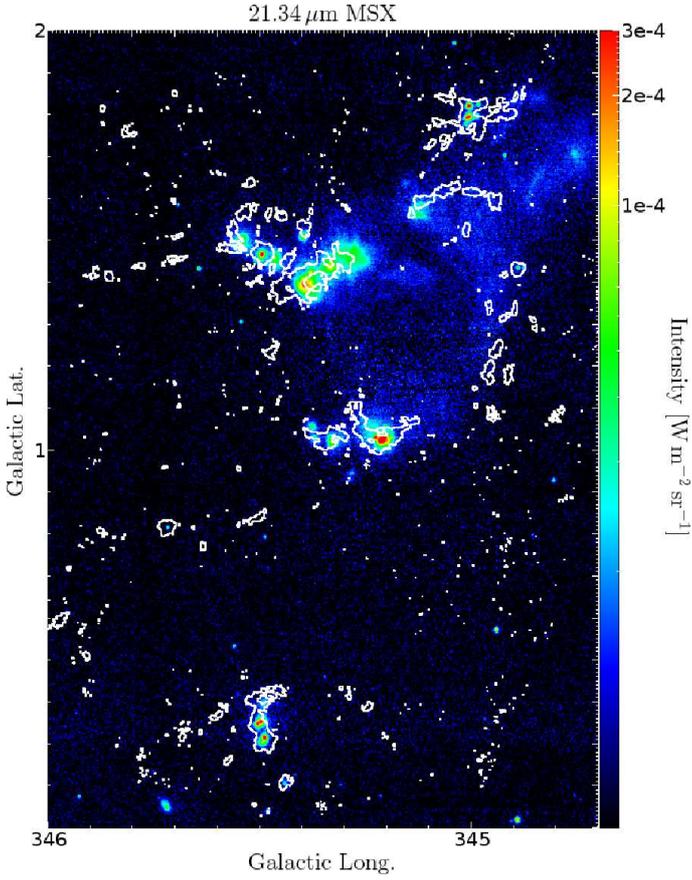}}
  \caption{ Image in 21.34\,$\mu$m from MSX observations toward GMC G345.5+1.0
   with contours of 1.2\,mm continuum emission at  three times rms, $\sim$0.06\,Jy\,beam$^{-1}$.}
\label{figureInfrared}
\end{figure}

Stars form in clumps, heating their surrounding dust, which 
  re-radiates at infrared wavelengths.  This is illustrated in
Fig. \ref{figureInfrared} that shows a strong spatial correlation
between the 1.2\,mm continuum emission and infrared  emission at
  21.34\,$\mu$m (from MSX observations).  To quantify the
correlation, we searched for infrared emission inside clump emission
areas, using  MSX\footnote{http://irsa.ipac.caltech.edu/} images at 8.28, 12.13,  14.65, and 21.34\,$\mu$m
and   SPITZER\footnote{http://irsa.ipac.caltech.edu/} (IRAC) images at
3.6, 4.5,  5.8, and 8.0\,$\mu$m.
We
find that $\sim$20\% of all clumps have an infrared counterpart
in all MSX and SPITZER bands (see Table \ref{tableClumps}).  
 The rest of the clumps, $\sim$80\%, are  not detected in all MSX and
SPITZER bands, 
particularly not in
12.13,  14.65, and 21.34\,$\mu$m.
Since 8.0\,$\mu$m MSX band and SPITZER IRAC bands are sensitive to 
the polycyclic aromatic hydrocarbon (PAH)
emission and the photospheric emission from stars (e.g. Chavarr\'{\i}a et al. 2008),
clump not detected in all MSX and
SPITZER bands  are considered to have
no counterpart at infrared wavelengths.
 Since both the  MSX and SPITZER
bands have sensitivity limits, the percentage of  detections is a lower
limit  to the number of clumps that are forming stars, and the
percentage of  failed  detections is an upper limit  to the number of clumps
that are not forming stars.

 Nine clumps are associated with six IRAS point sources classified as MSFRs
with luminosities $\gtrsim$10$^3$\,L$_\odot$ (see
Sect. \ref{sectionDustIRAS} and Table \ref{tableSED}).  As
Fig. \ref{figureSize-Mass} shows, these clumps have densities of
$\sim$10$^5$\,cm$^{-3}$, suggesting that there is a threshold density
 above which massive stars can form.
 These  values are consistent with the typical
density of clumps associated with MSFRs ($\sim$10$^5$\,cm$^{-3}$;
Fa\'undez et al. 2004).

As Fig. \ref{figureSize-Mass} shows, clumps 
 that emit detectable
infrared
emission tend to be more massive than 
remaining  clumps.  Clumps
without infrared emission (cold or starless clumps) have a  mean mass
of 21\,M$_\odot$, and clumps with an infrared counterpart have a
mean mass of 2.1$\times$10$^2$\,M$_\odot$.  Furthermore, all clumps
with masses  higher than $\sim$200\,M$_\odot$ have an infrared
counterpart.

 MSX point sources associated with clumps within the GMC G345.5+1.0 have
mid-infrared colors $S_{21}$/$S_{8}$ from  0.9 to 30, $S_{14}$/$S_{8}$
from 0.4 to 8, and $S_{12}$/$S_{8}$ from 0.7 to 4, where $S_{8}$,
$S_{12}$,  $S_{14}$, and $S_{21}$  are the flux densities at 8.28, 12.13,
14.65, and 21.34 $\mu$m, respectively.  These ratios cover those of massive
young stellar objects (MYSOs; Lumsden et al. 2002): $S_{21}/S_{8}>2$
and $S_{21}>S_{14}>S_{8}$. About 7\% of the clumps  contain  MSX sources 
that satisfy this criterion, and these clumps have masses 
$\gtrsim$36\,M$_\odot$ (see Fig. \ref{figureSize-Mass}).

 The existence of clumps with and without infrared emission suggests
that clumps in the GMC are  at different evolutionary stages.  Cold
clumps have masses between 3.0 and 1.9$\times$10$^2$\,M$_\odot$, where
the most massive ones are possible progenitors of MSFRs.  
For example, we estimate that the least massive clump associated with a MYSO
has  a mass of $\sim$36\,M$_\odot$, 
and  we  identify seven cold clumps with
densities $\gtrsim$10$^5$\,cm$^{-3}$ and 
masses $\gtrsim$36\,M$_\odot$, which will eventually collapse to form
high-mass stars.

Do clumps form single stars?  One way to assess  this
is to compare the slope of the clump mass distribution with that of
the stellar initial mass function (IMF) (e.g. Motte et al. 1998; Lada
et al. 2007).
Equal slopes  would indicate
that the origin of the stellar IMF has its direct roots in the origin
of the clump mass distribution.
The spectral mass index $\alpha$ of
the clump mass distribution determined here is consistent with  that of other
investigations (e.g. Mu\~noz et al. 2007), 
 but differs 
from
that estimated for the stellar IMF of the solar neighborhood for
stellar masses  higher than 0.5\,M$_\odot$ ($\alpha$$\sim$2.35;
 e.g. Kroupa 2002).  This suggests that the detected
clumps do not directly form stars, and other processes are necessary to
determine the stellar initial masses,  such as the fragmentation of
clumps, mainly of the most massive ones. 
Figure \ref{figureCMD} compares the IMF spectral mass 
index with the clump mass distribution.

\subsection{Dust properties of massive-star forming regions  associated with  clumps and IRAS point sources}
\label{sectionDustIRAS}

Regions of massive star formation are embedded in massive clumps, and
the intense Lyman flux produced by them heats the surrounding dust,
which re-emits mainly  at far infrared wavelengths with characteristic
colors.  To study the physical properties of dust in these regions, we
examine the spectral energy distributions (SEDs) of MSFRs associated
with IRAS point sources and clumps detected here, assuming that their
emissions are from dust.

\begin{figure*}[!h]
  \resizebox{\hsize}{!}{\includegraphics{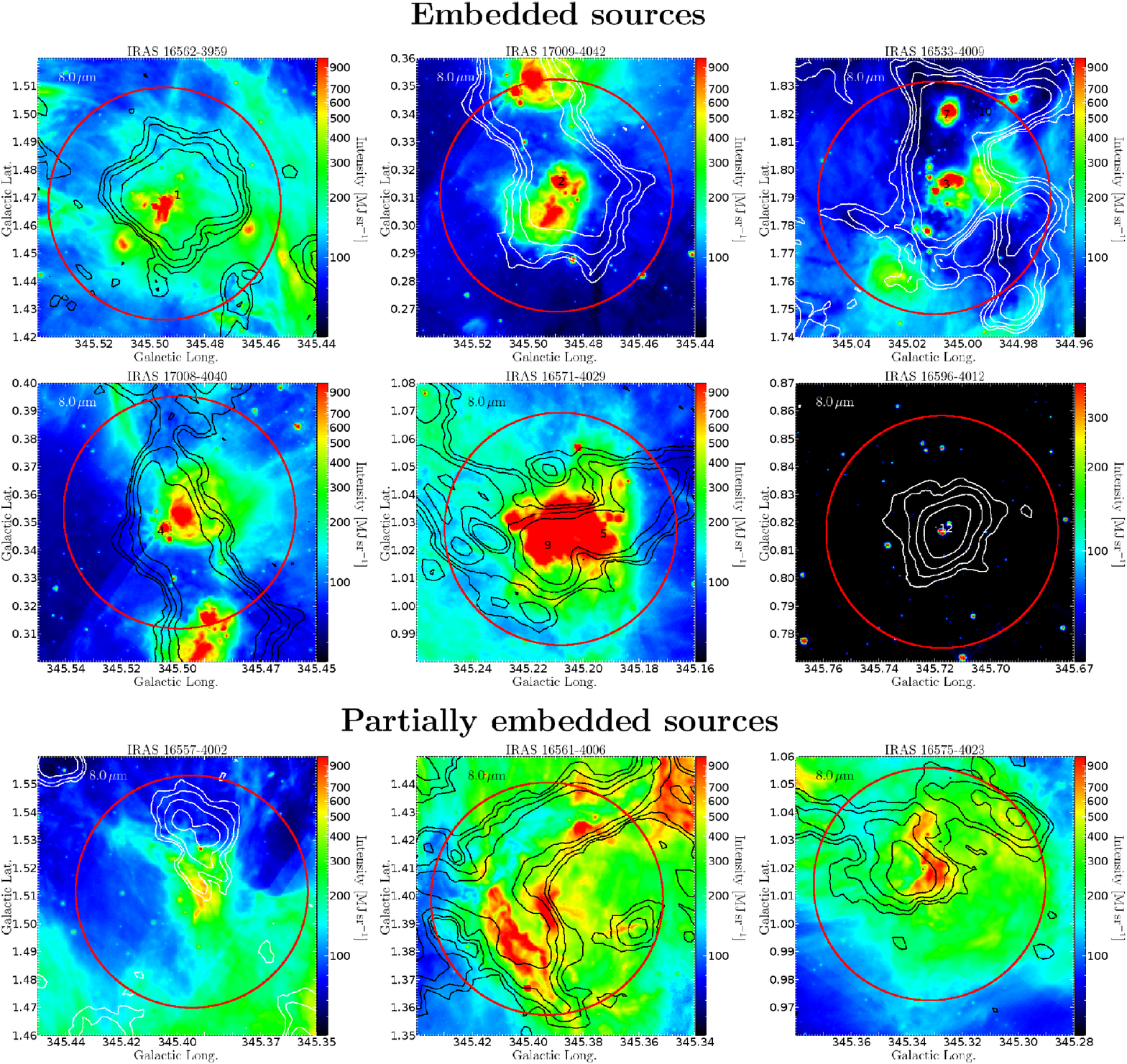}}

  \caption{Images of 8.0\,$\mu$m emission (SPITZER data) toward clumps
    detected in 1.2\,mm continuum emission and associated with IRAS
    point sources.  Contours represent 1.2\,mm continuum emission at
    0.06, 0.12,  0.24, and 0.48\,Jy\,beam$^{-1}$ (rms is 0.02\,Jy\,beam$^{-1}$).    IRAS source
    names are  given at the top of each image, and  clump numbers are indicated at the peak
    of 1.2\,mm continuum emission. Red circles are  centered
    on the coordinates of IRAS point sources, with diameters of 5$'$
    (an approximation of the angular resolution of IRAS observations
    at 100\,$\mu$m).}
\label{figureMSFRs}
\end{figure*}

Within the GMC, there are eight MSFRs associated with IRAS-CS sources
(Table \ref{tableCS}).  
 We added one more source, IRAS
  16533-4009, which is embedded in 1.2\,mm continuum emission, has a
  high luminosity, $\sim$9$\times$10$^4$ L$_\odot$, and  increasing IRAS
  flux densities from 12 to  100\,$\mu$m; however it does not satisfy
  the  far-infrared color criterion defined by Wood \& Churchwell
  (1989), since its flux density in 25\,$\mu$m is an upper limit.
Figure \ref{figureMSFRs} shows 8.0\,$\mu$m
images from SPITZER data with contours of 1.2\,mm continuum emission
for all these sources.
For  six of these objects, the SPITZER emission is embedded within
1.2\,mm continuum emission.  For  more reliable  estimates in our SED study,
we only consider these  six sources: 
IRAS 16533-4009,
IRAS 16562-3959,
IRAS 16571-4029,
IRAS 16596-4012,
 IRAS 17008-4040, and
IRAS 17009-4042.

\begin{figure*}[h]
\resizebox{\hsize}{!}{
\includegraphics[width=230pt]{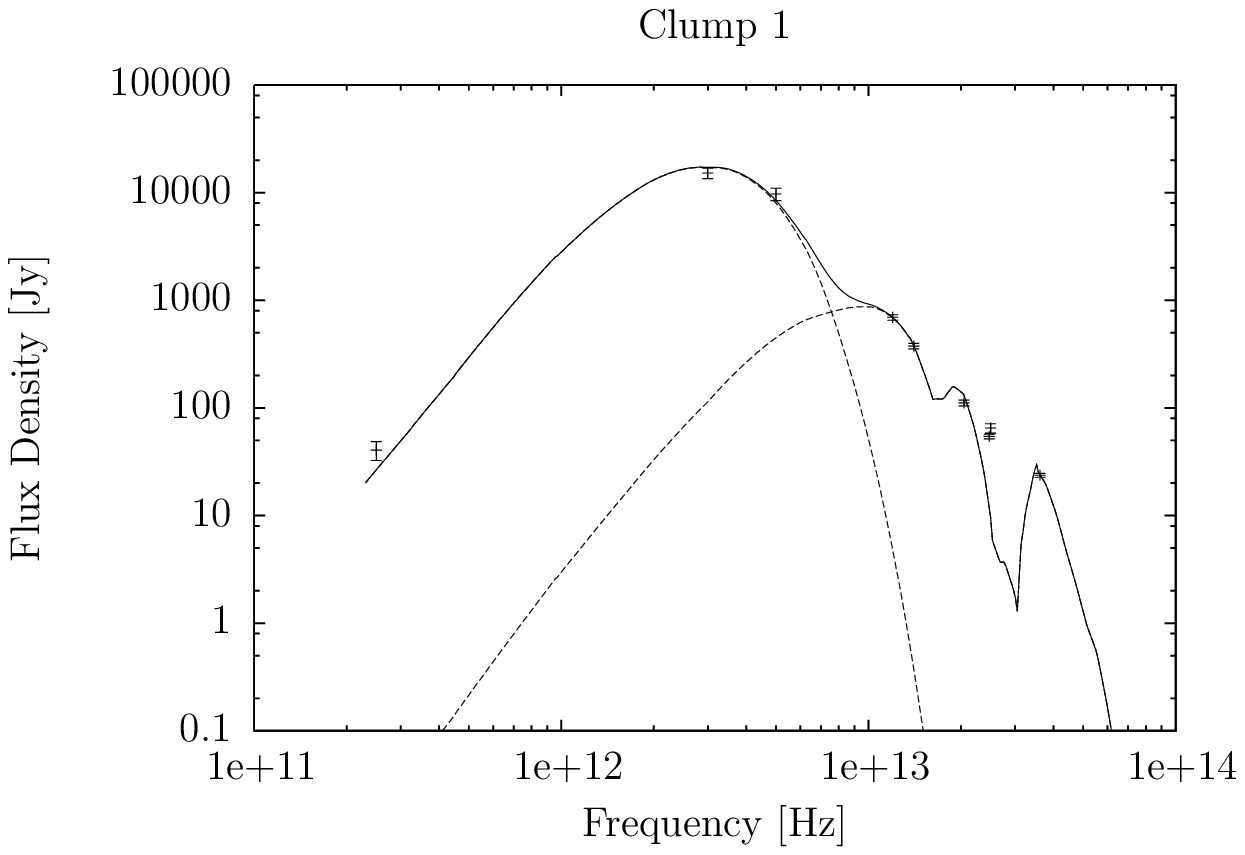}
\includegraphics[width=230pt]{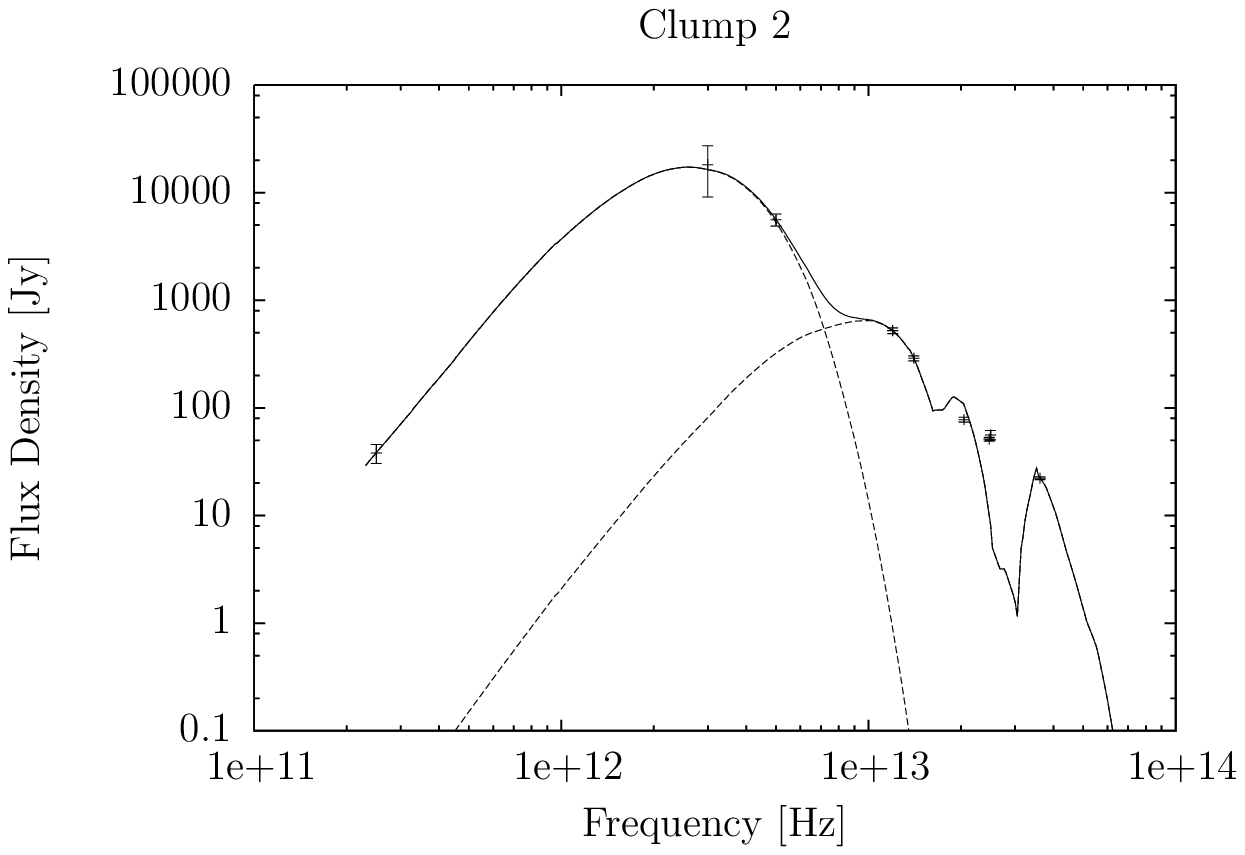}}
\resizebox{\hsize}{!}{
\includegraphics[width=230pt]{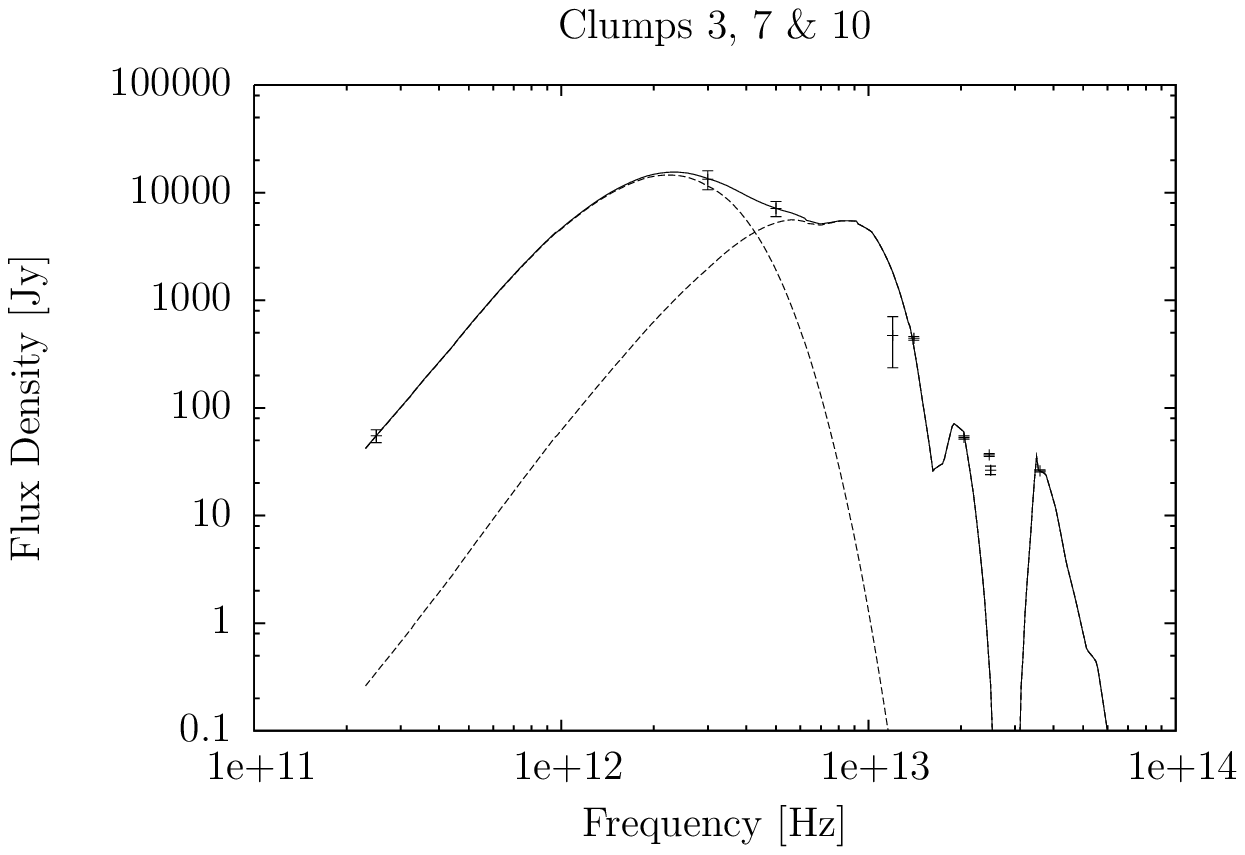}
\includegraphics[width=230pt]{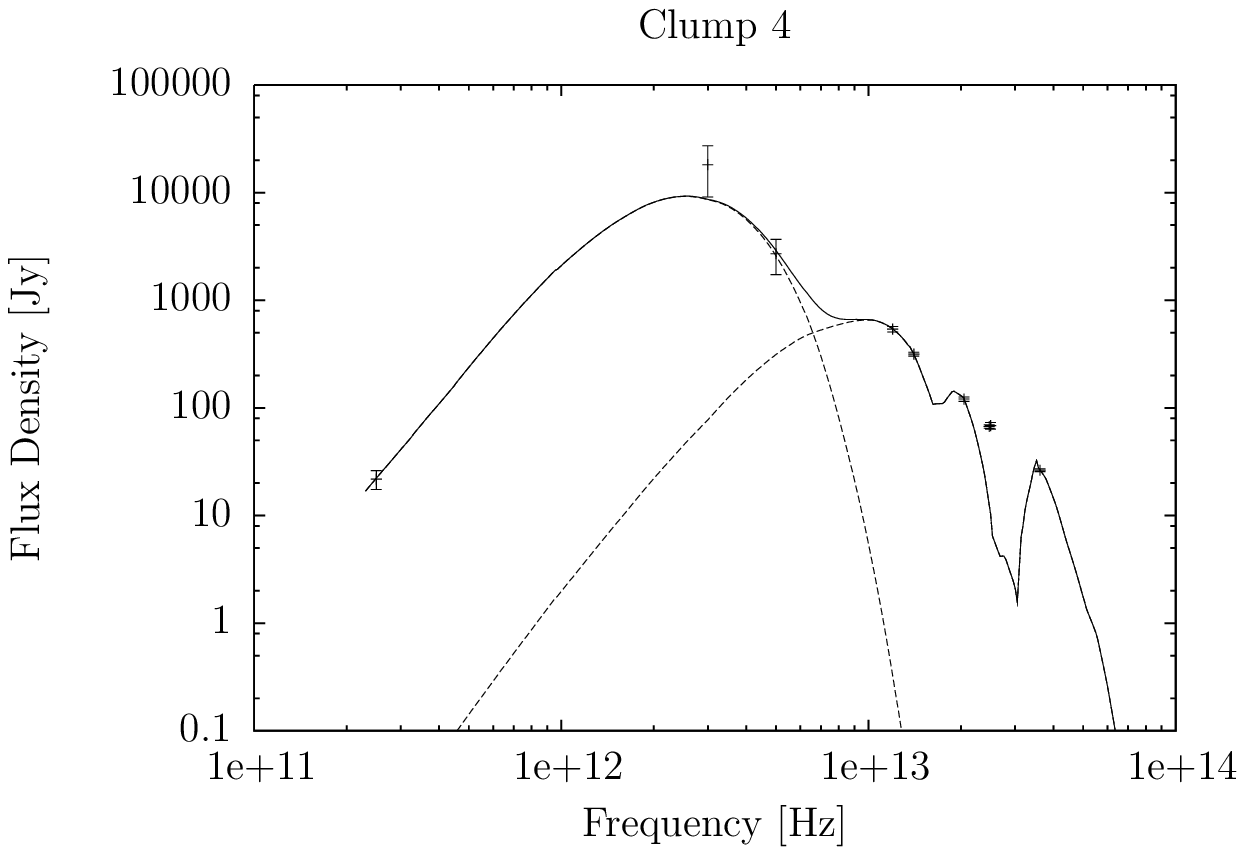}}
\resizebox{\hsize}{!}{
\includegraphics[width=230pt]{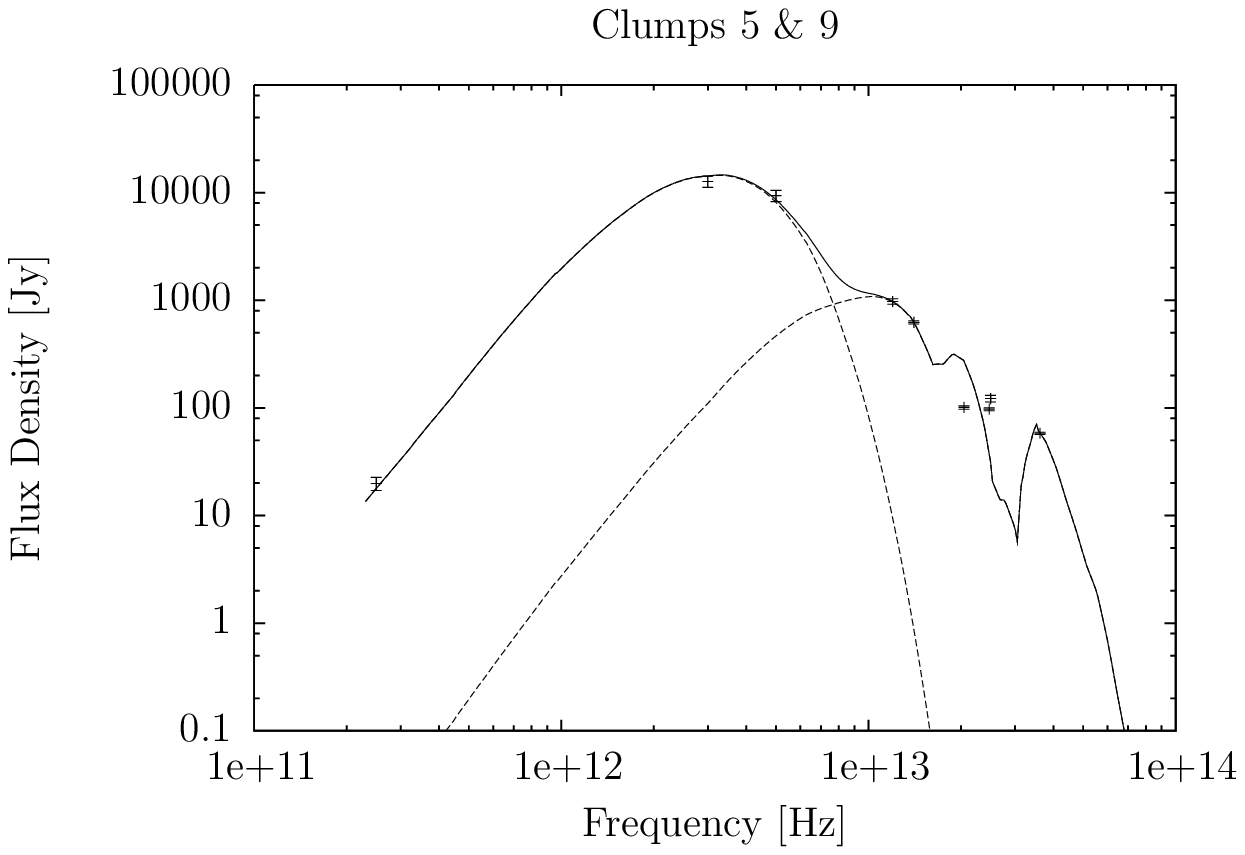}
\includegraphics[width=230pt]{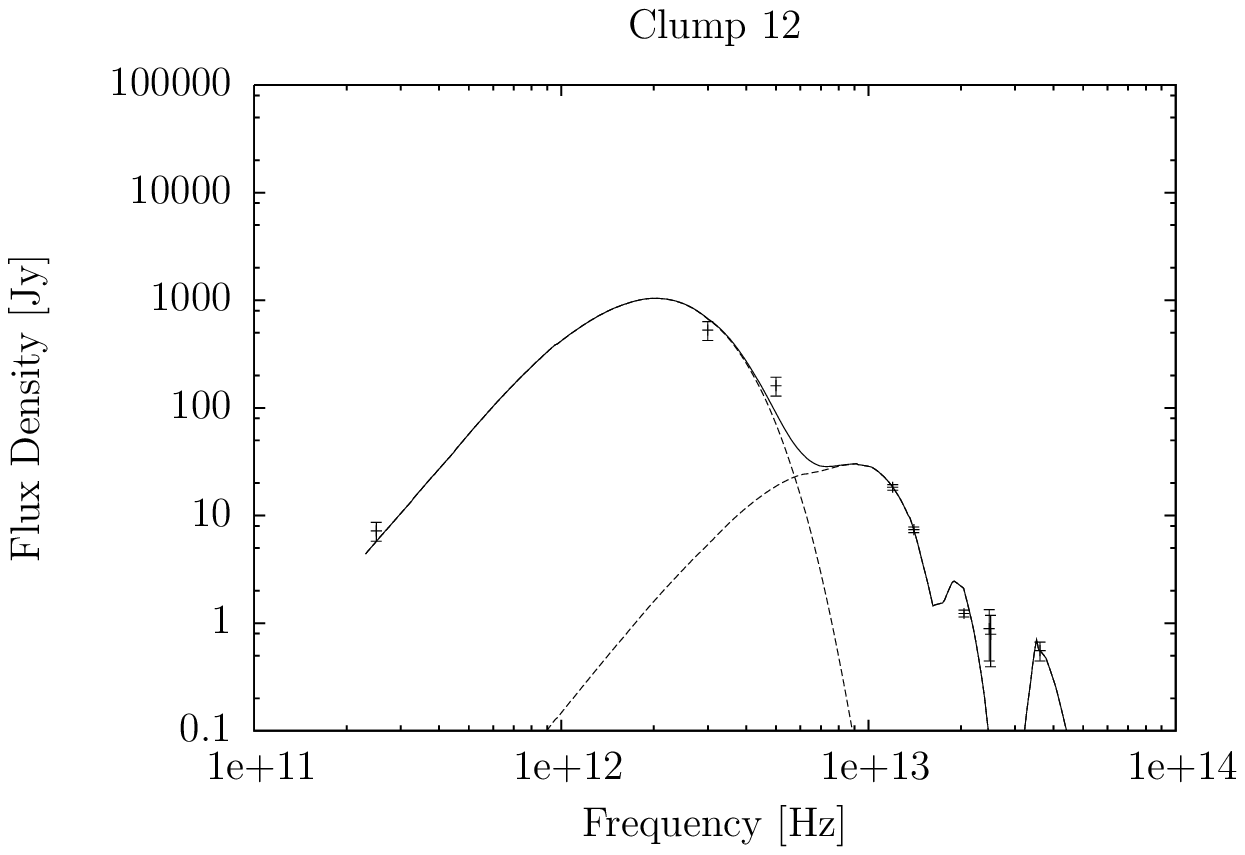}}
\caption{ The SEDs of massive-star forming
  regions associated with massive clumps detected in 1.2\,mm continuum
  emission; top labels show names of the clumps.  Dots with error bars
  are flux densities estimated from SIMBA,  IRAS, and MSX observations.
  Each SED is modeled with two dust components at different
  temperatures (physical parameters for each model are in Table
  \ref{tableSED});  drawn lines are the total flux density of the
  two dust components, and dashed lines are  the contributions of each dust
  component.}
\label{figureSEDs}
\end{figure*}

 Figure \ref{figureSEDs} displays the six SEDs constructed using our
observations in 1.2\,mm continuum emission, infrared data at 12, 25,
 60, and 100\,$\mu$m from  the IRAS Point Source Catalog (version 2.0), and
at 8.3, 12.1, 14.7 and 21.3\,$\mu$m from  the MSX Point Source Catalog
(version 2.3).   Because  observations were  performed using different beam
  sizes, for IRAS $\sim$300$''$, for MSX  $\sim$20$''$, and for 1.2\,mm
  24$''$, we  consider all emission  within the IRAS beam;
thus,  two
SEDs  are associated with more than one 1.2\,mm clump.

\begin{figure}[!h]
  \resizebox{\hsize}{!}{\includegraphics{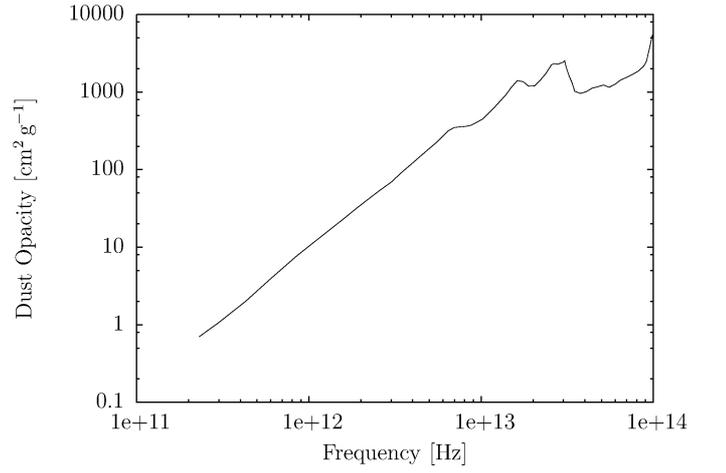}}
  \caption{ Dust opacity spectrum utilized in the SED models.  
    It was estimated by   Ossenkopf \& Henning
    (1994)$^a$,
    assuming a  Mathis-Rumpl-Nordsieck  initial size distribution
    with thin ice mantles and 10$^5$\,yr of coagulation  at a gas density
    of 10$^5$\,cm$^{-3}$.}
    $^a$   http://vizier.u-strasbg.fr/viz-bin/VizieR, J/A+A/291/943/table1
  \label{figureDustOpacity}
\end{figure}

The SEDs of MSFRs can be modeled as several dust components at different
temperatures (e.g. Fa\'undez et al. 2004; Morales et al. 2009).
Because of the  shape of the  six SEDs, we model them as two dust
components at different temperatures, cold and warm components,
including the absorption of the radiation by assuming that the
warm component is embedded in the cold one.  The total flux density,
$S_\nu^{total}$, at frequency $\nu$ is approximated  by
$$S_\nu^{total} \sim S_\nu^{cold}+ S_\nu^{warm},$$ where
$$S_\nu^{cold}= \Omega^{cold}B_\nu (T_{dust}^{cold})
(1-\exp(-\tau_\nu^{cold}))$$ 
and 
$$S_\nu^{warm}= \Omega^{warm}B_\nu
(T_{dust}^{warm})
(1-\exp(-\tau_\nu^{warm}))\exp(-\tau_{\nu}^{cold}/2).$$
 The parameters $S_\nu^{cold}$, $\Omega^{cold}$,  $T_{dust}^{cold}$, and
$\tau_{\nu}^{cold}$ are the flux density, the solid angle, the dust
 temperature, and the optical depth of the cold component,  respectively,
and 
$S_\nu^{warm}$, $\Omega^{warm}$,  $T_{dust}^{warm}$, and
$\tau_{\nu}^{warm}$ are the flux density, the solid angle, the dust
 temperature, and the optical depth of the warm component, respectively.
In addition, $B_\nu(T_{dust}^{cold})$ and $B_\nu(T_{dust}^{warm})$ are the Planck
function at dust temperatures $T_{dust}^{cold}$ and $T_{dust}^{warm}$,
respectively.  For both components, $\Omega$ can be expressed as
$$\Omega=\pi\left(\theta/2 \right)^2,$$ where $\theta$ is the angular
diameter.
The optical depths are given by  (e.g. Evans 1999)
$$\tau^{cold}_\nu= N^{cold}_{dust}\,k_\nu\ \, \ \,\ and\ \, \ \,
\ \tau^{warm}_\nu= N^{warm}_{dust}\,k_\nu,$$ where $N^{cold}_{dust}$
and $N^{warm}_{dust}$ are the dust column densities in g\,cm$^{-2}$
for the cold and warm components, and $k_\nu$ is the dust opacity.  We
use dust opacities estimated by Ossenkopf \& Henning (1994) for
protostellar cores. They computed opacities considering the
Mathis-Rumpl-Nordsieck 
(MRN) distribution for the diffuse interstellar
medium (Draine \& Lee 1984) as the initial size distribution for dust,
without and with ice (thin and thick), and without and with
coagulation (after 10$^5$\,years for densities between
10$^5$-10$^8$\,cm$^{-3}$).   In the case of  regions with  ice
depletion produced by   the heating of  central sources, they  recommended
opacities for the model with thin ice mantles and coagulation for a
density of 10$^5$\,cm$^{-3}$.   These dust opacities are shown in
Fig. \ref{figureDustOpacity}, for
frequencies between $\sim$2.3$\times$10$^{11}$ and
10$^{14}$\,Hz.

Thus, in our SED model, each dust component has three values to fit of
$T_{dust}$,  $\theta$, and $N_{dust}$.  However, the dust column density
of the warm component, $N_{dust}^{warm}$, is difficult to estimate,
because it is more sensitive to the emission in the Rayleigh-Jeans
part of the spectrum ($h$$\nu$$<<$$k$\,$T_{warm}$), where the emission
is dominated by the cold component.  To overcome this problem, we
 assume that the two components have equal densities, thus
$$N_{dust}^{warm}\sim\frac{\theta_{dust}^{warm}}
{\theta_{dust}^{cold}}N_{dust}^{cold}.$$ 
Given the simplicity of
  the SED model and the poor sensitivity of the data to $N_{dust}^{warm}$, 
  a more realistic density distribution is unnecessary.

 To enable a more reliable
comparison, angular diameters are converted
into spatial diameters, and dust column densities to gas column
densities.  Thus 
$$D_{dust}^{cold} =  d\,\theta^{cold}_{dust},$$
$$D_{dust}^{hot} =  d\,\theta^{warm}_{dust},$$
and 
$$N_{gas}^{cold}=\frac{N_{dust}^{cold} R_{gd}}{\mu\,m_{H}},$$
 where $\theta^{cold}_{dust}$ and $\theta^{warm}_{dust}$ are in 
radians.
In this way, our model has  five variables: dust temperatures and sizes
for the two components, and  gas column density for the cold
component.

Table \ref{tableSED} and Fig. \ref{figureSEDs} display the results of
the fits.   The mean dust temperatures  of each component are  28$\pm$5\,K
 (cold) and 200$\pm$10\,K (warm).   The sizes and column densities of the
cold  component agree with those estimated by 1.2\,mm
continuum: sizes  vary by a factor of   0.7-1.5 and  column densities vary by a
factor of  0.5-3.  Estimates of luminosities, from  the integration of
SED models, are $>$10$^3$\,L$_\odot$.  Given the sizes and column
densities, the total mass is dominated by the cold component
($\sim$99\% of the total mass), and is similar to that  estimated from the 
1.2\,mm continuum emission, varying by a factor of 0.8-1.6.

Dust characteristics of clumps associated with MSFRs estimated in this
paper are consistent with previous works (e.g. Fa\'undez et al. 2004;
Molinari et al. 2000; Molinari et al. 2008), where the cold dust
temperature in regions of massive star formation is $\sim$30\,K.

 The  SED models have a discrepancy with data in $\sim$12\,$\mu$m bands
  (see Fig. \ref{figureSEDs}), which can be produced by not considering
  the PAH  emission in the SED models (e.g. van Dishoeck 2004), 
  or by
  an excess in the dust opacities utilized at these wavelengths,
  affecting the modeled radiation from the hot component  that is
  absorbed by the cold one.

\begin{figure*}[h]
\resizebox{\hsize}{!}{
\includegraphics[width=230pt]{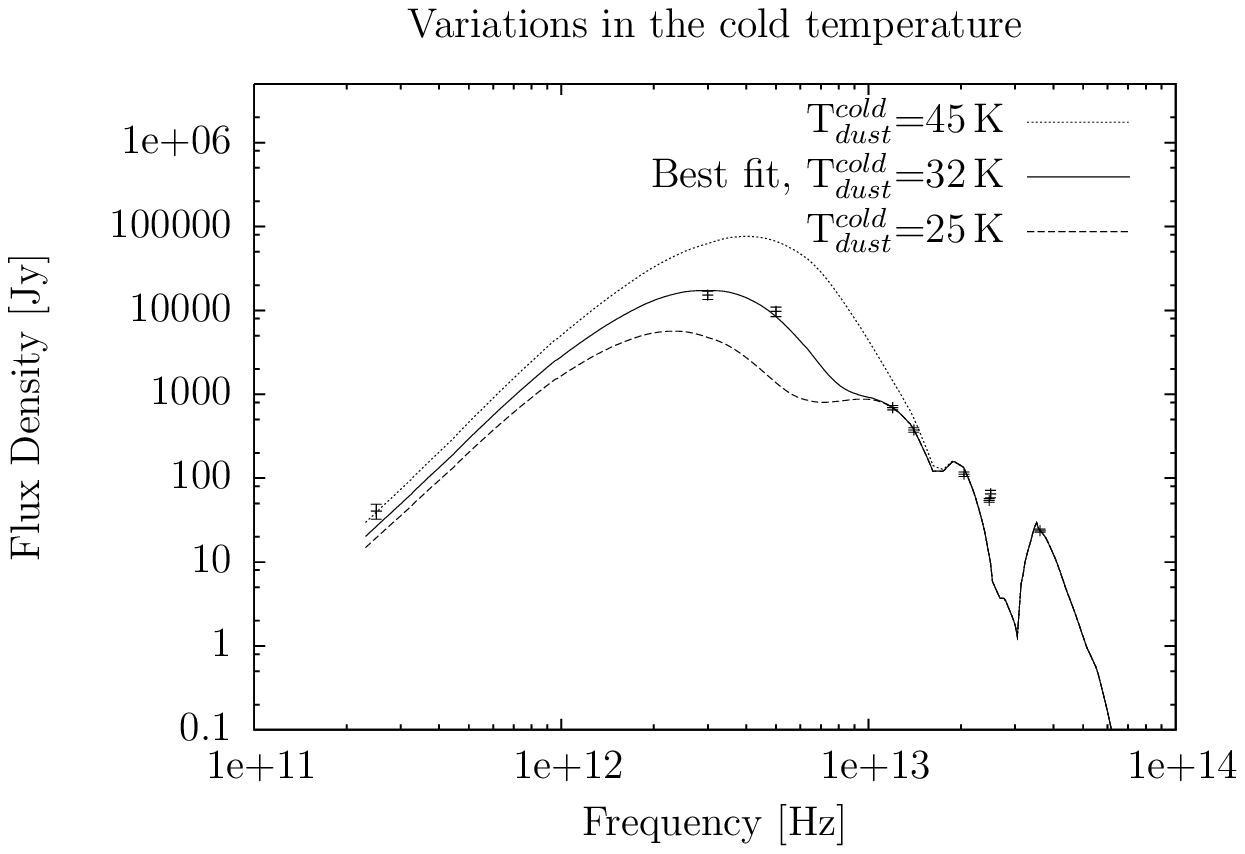}
\includegraphics[width=230pt]{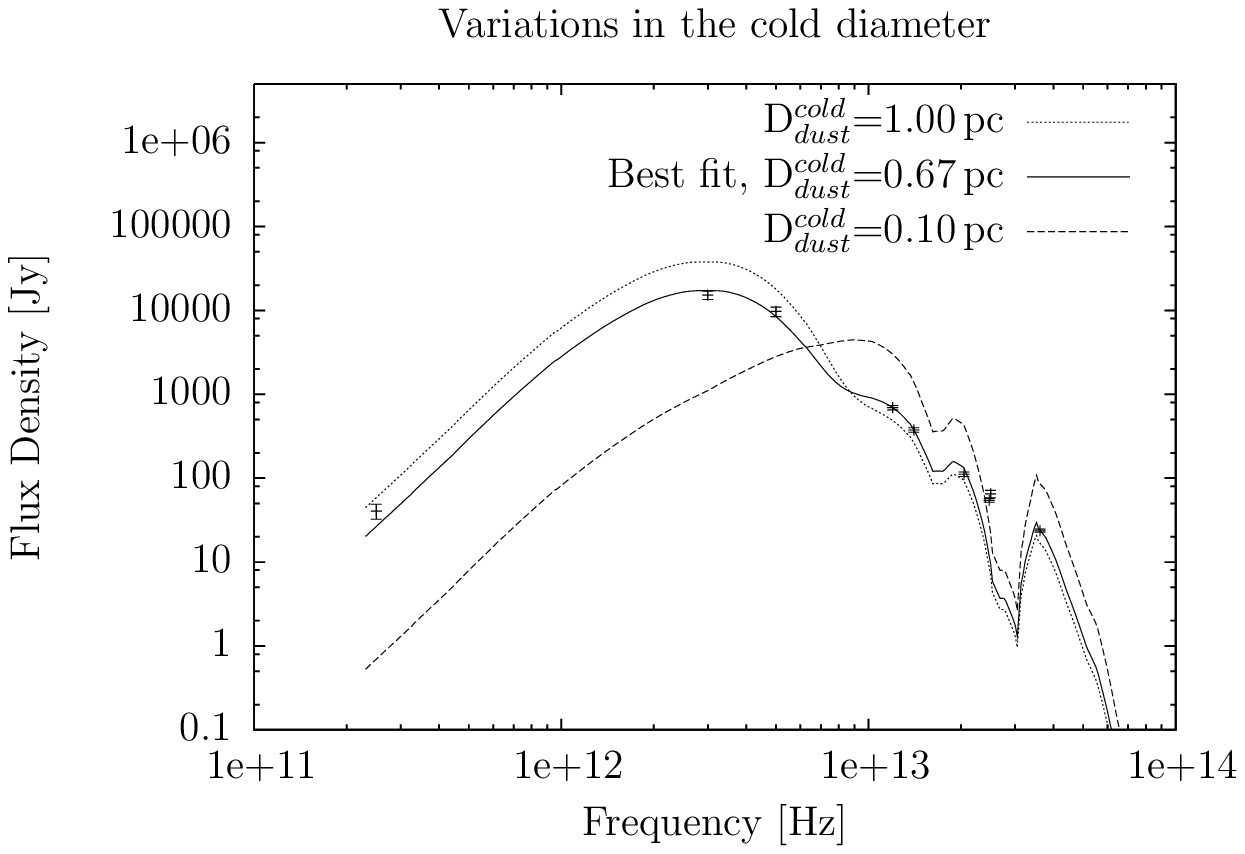}}
\resizebox{\hsize}{!}{
\includegraphics[width=230pt]{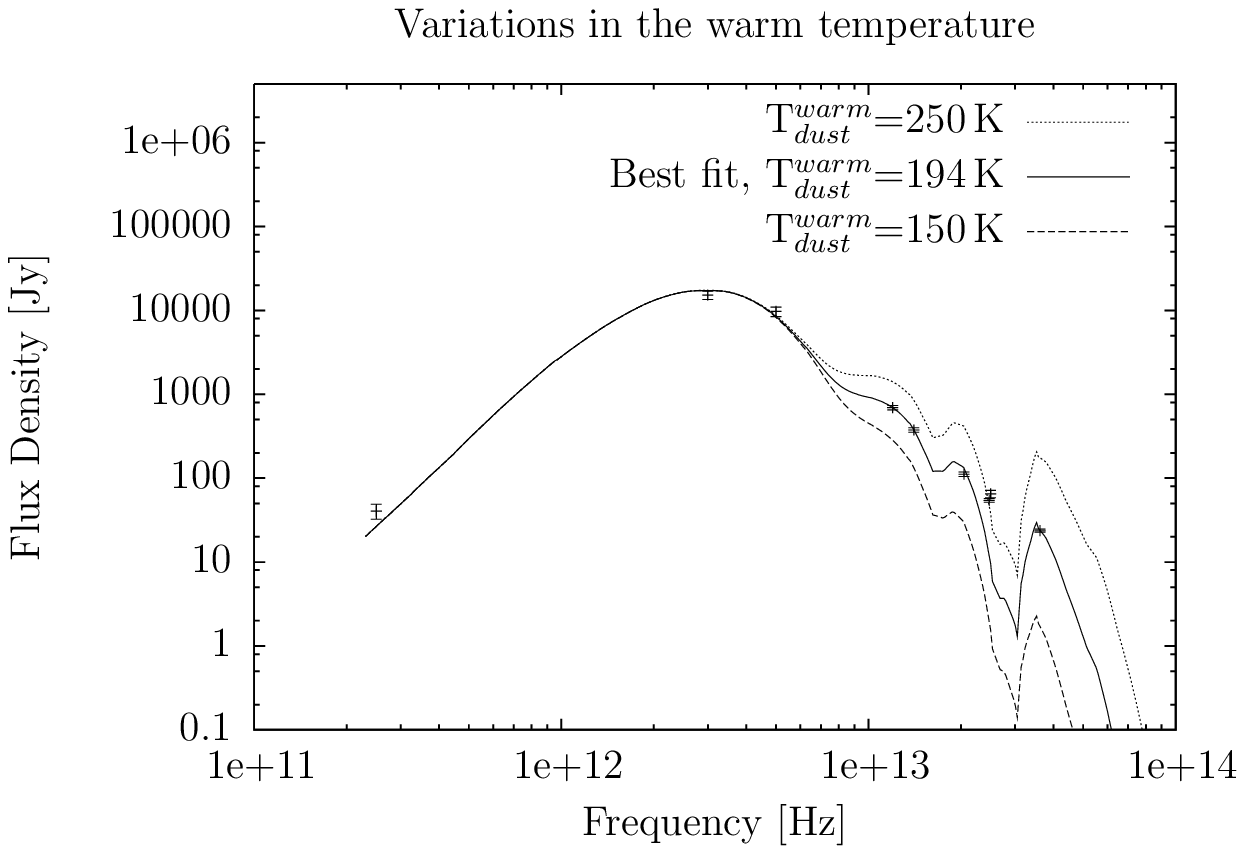}
\includegraphics[width=230pt]{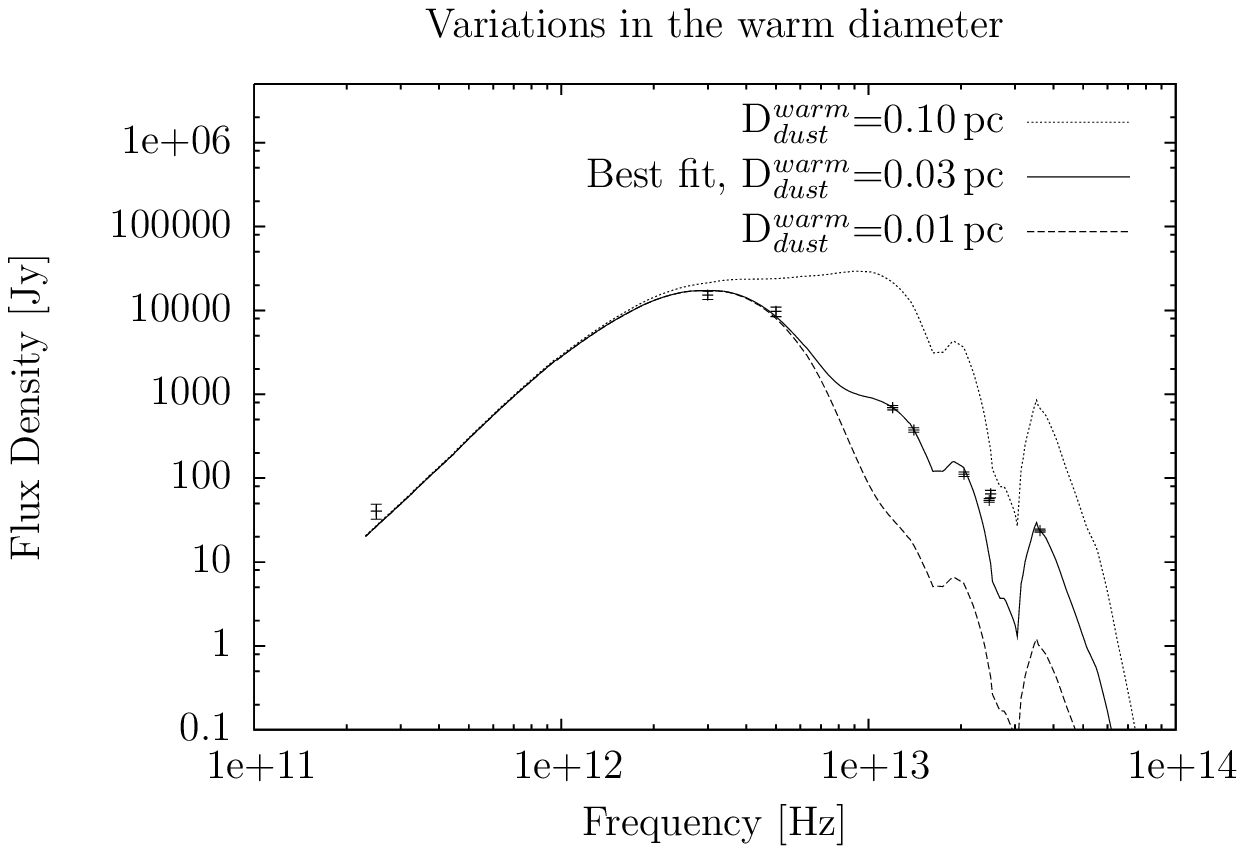}}
\resizebox{260pt}{!}{\includegraphics[width=230pt]{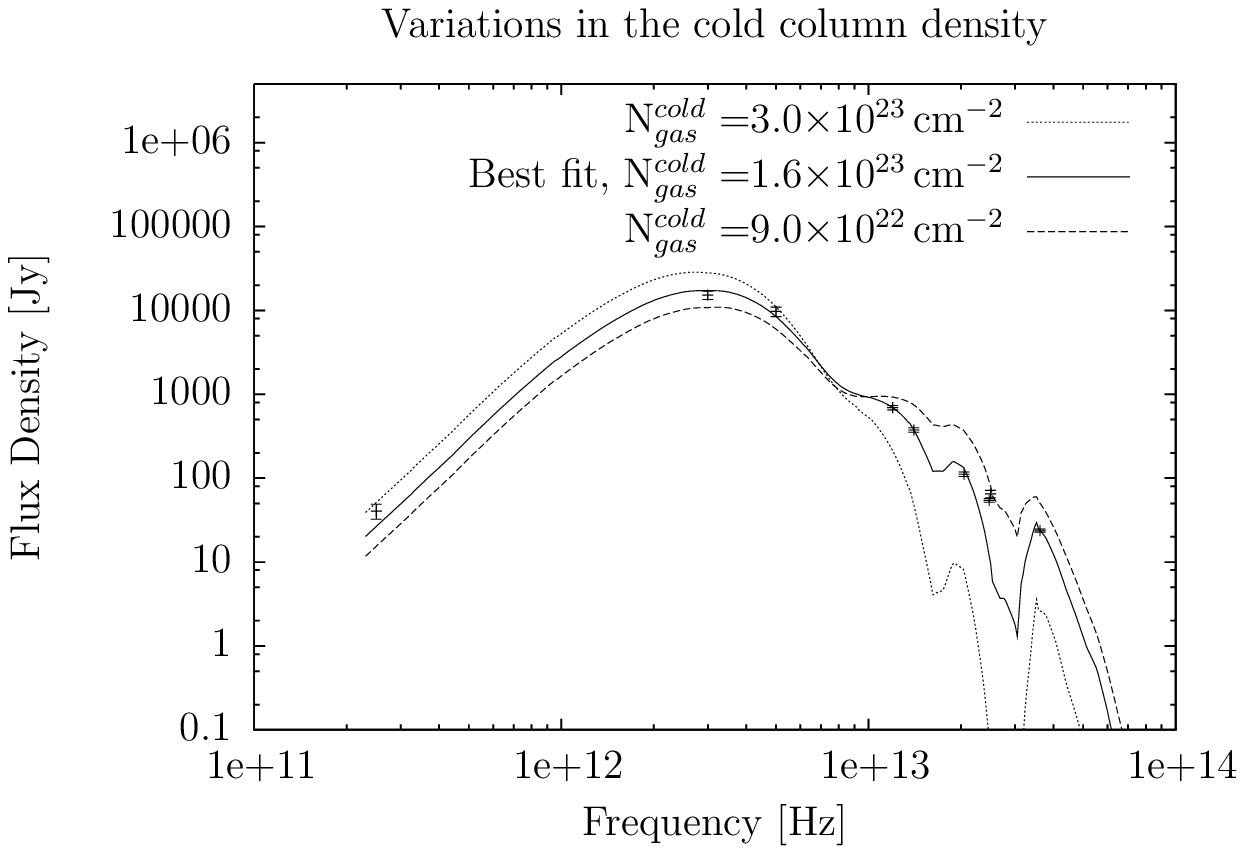}}
\caption{ Dependence of the SED model  on variations  in the fitted parameters.
Plots show the SED  for clump 1 with the  best-fit model (see Table \ref{tableSED}) and 
variations  in each parameter, for 
$T_{dust}^{cold}$, $D_{dust}^{cold}$,
$T_{dust}^{warm}$, $D_{dust}^{warm}$ and $N_{gas}^{cold}$, 
 when increasing and decreasing  the  best-fit value.}
\label{figureSEDVariation}
\end{figure*}

The dependence of SED models  on variations in the fitted parameters is
shown in Fig. \ref{figureSEDVariation}, which displays the SED for
clump 1, or IRAS 16562-3959, with the  best-fit model and models
with  variations in the  fitted parameters $T_{dust}^{cold}$,
$D_{dust}^{cold}$, $T_{dust}^{warm}$,  $D_{dust}^{warm}$, and
$N_{gas}^{cold}$.  For each dust component, variations  in its
temperature  affect both its peak emission frequency and its luminosity,
 both 
 values increasing  as the temperature increases.  Variations  in either
$D_{dust}^{cold}$ or $N_{gas}^{cold}$ affect the luminosities of the
two components; when $D_{dust}^{cold}$ or $N_{gas}^{cold}$ increases,
the luminosity of the cold dust component increases,  whereas radiation
absorption of the warm component also increases, reducing its
luminosity.  When $D_{dust}^{warm}$ increases, the luminosity also
increases.

\section{Conclusions}
\label{sectionConclusions}

We  have robustly detected the whole of the  GMC G345.5+1.0 in 1.2\,mm continuum emission at
a spatial resolution of 0.2\,pc, and conclude that:
\begin{itemize}
\item The GMC is fragmented. We  have identified  201 clumps, which have
   beam-corrected diameters  between 0.2 and 0.6\,pc, masses between 3.0 and
  1.3$\times$10$^3$\,M$_\odot$, and densities between 5$\times$10$^3$
  and 4$\times$10$^5$\,cm$^{-3}$.\\
\item The total mass of the clumps is $\sim$1.2$\times$10$^4$\,M$_\odot$,
  and  after comparing with the total mass of the GMC of
  $\sim$6.5$\times$10$^5$\,M$_\odot$,  we inferred that the efficiency in forming these
  clumps is $\sim$0.02.\\
\item The clump mass distribution is well-fitted by a power law
  dN/dM$\propto$M$^{-\alpha}$, where the spectral mass index $\alpha$
  is 1.7$\pm$0.1. The total mass is dominated by
  massive clumps, but the population is dominated by clumps with low masses.\\
\item  The spectral mass index of the clump mass distribution is
  different from that of the stellar IMF.  Thus  our detected clumps  are 
  probably not  the direct progenitors of single stars.  \\
\item Comparing with MSX and  SPITZER (IRAC-bands) observations,
  20\% of the clumps have an infrared counterpart in all MSX and
  SPITZER bands.   The  remaining clumps, $\sim$80\%, are considered to
  have no counterpart  at infrared wavelengths.  The percentage of
  detection is a lower limit  to the number of  clumps forming
  stars, while the percentage of no  detections is an upper limit  to
  the number of clumps that are not forming stars.\\
\item Regions of  massive-star formation within the cloud,
  associated with IRAS point sources, have  SEDs that can be
  modeled with two dust components at  different  mean
  temperatures of 28$\pm$5 and 200$\pm$10\,K.\\
\end{itemize}

\begin{acknowledgements}

  C.L.  acknowledges partial support from the GEMINI-CONICYT FUND,
  project number 32070020, and ESO-University of Chile Student
  Fellowship.  This work was supported by the Chilean Center for
  Astrophysics FONDAP N$^\circ$ 15010003 and by Center of Excellence
  in Astrophysics and Associated Technologies PFB 06.

\end{acknowledgements}

\clearpage 
\onecolumn

\newcommand{\tableSEDL}
{IRAS & Clumps  &\multicolumn{3}{c}{Cold} & \multicolumn{3}{c}{Warm} &  Cold mass & Luminosity \\ 
      &&  T$_{dust}$ & D$_{dust}$ & N$_{gas}$ & T$_{dust}$ &  D$_{dust}$ & N$_{gas}$&&\\
      &&     K & pc & 10$^{23}$\,cm$^{-2}$& K & pc &10$^{22}$\,cm$^{-2}$&10$^3$\,M$_\odot$&10$^4$\,L$_\odot$\\}

\begin{center}
\setlongtables
\begin{longtable}{cccccccccc}
 \caption{SED models for MSFRs associated with IRAS point sources and
   massive clumps detected in 1.2\,mm continuum emission.  Each model
    consists of two dust components with equal densities at
   different temperatures (cold and warm components).  Column 1 shows
   names of IRAS point sources; Column 2, names of clumps;  in
     Columns 3 to  8, we show the fitted physical parameters: dust
     temperature,  diameter, and column density, respectively, for the
     cold (Cols. 3-5) and warm (Cols. 6-8) components; Column 9,
   masses of the cold component; and Column 10, total luminosities.}\\
\label{tableSED}\\
\hline\hline\\
\tableSEDL
\hline\\
\endfirsthead

\multicolumn{10}{c}{\bfseries \tablename\ \thetable{} -- continued from previous page} \\
\hline\hline\\
\tableSEDL
\hline\\
\endhead

\hline\\ 
\multicolumn{10}{c}{{Continued on next page}} \\
\endfoot

\hline \\

\endlastfoot

16562-3959 & 1&32&0.67&1.6&194&0.03&0.69&1.0&5.3\\

\hline
\\
17009-4042 & 2&29&0.86&1.6&202&0.03&0.51&1.7&4.7\\

\hline
\\
16533-4009 &  3, 7, 10&25&0.79&3.2&205&0.07&2.7&2.8&9.4\\

\hline
\\
17008-4040 & 4&28&0.68&1.5&205&0.03&0.57&1.0&2.9\\

\hline
\\
16571-4029 &  5, 9&34&0.58&1.3&209&0.03&0.65&0.64&5.3\\

\hline
\\
16596-4012 & 12&22&0.34&2.2&203&0.008&0.49&0.36&0.22\\

\\

\end{longtable}
\end{center}

\clearpage 

\clearpage

\newcommand{\tableClumpsKey}
{
Name& \multicolumn{2}{c}{Galactic coord.}& S$_{1.2mm}$ &Diameter & Mass & Density & Column density & Infrared\\
    &  long.   & lat.             & Jy        &pc&  M$_\odot$&   cm$^{-3}$ &   cm$^{-2}$ & counterpart\\\\
}

\onltab{5}{
\begin{center}
\setlongtables
\begin{longtable}{ccc cccc cc}

\caption{Properties of identified clumps in the GMC G345.5+1.0. Column
  1 gives clump names; columns 2 and 3,  Galactic coordinates of peaks
  in 1.2 mm continuum emission; column 4, 1.2 mm flux densities;
  column 5, diameters;  column 6,
  masses; column 7,  densities; column 8, column densities; and column
  9, if clumps are detected  in all infrared MSX and SPITZER-IRAC bands (``Y'') or not  (``N'').
  The densities and the column densities are computed assuming a mean molecular
weight of $\mu$=2.29.}\\
\label{tableClumps}\\
\hline\hline\\
\tableClumpsKey
\hline\\
\endfirsthead

\multicolumn{9}{c}%
{{\bfseries \tablename\ \thetable{} -- continued from previous page}} \\
\hline\hline\\
\tableClumpsKey
\hline\\
\endhead

\hline\\ 
\multicolumn{9}{c}{{Continued on next page}} \\
\endfoot

\hline \\

\endlastfoot
1&345.490&1.471&40&0.5&1.3$\times$10$^3$&2.9$\times$10$^5$&3.2$\times$10$^{23}$&Y\\
2&345.488&0.316&38&0.6&1.3$\times$10$^3$&2.3$\times$10$^5$&2.7$\times$10$^{23}$&Y\\
3&345.006&1.795&32&0.5&1.1$\times$10$^3$&3.9$\times$10$^5$&3.6$\times$10$^{23}$&Y\\
4&345.506&0.347&22&0.6&7.2$\times$10$^2$&1.4$\times$10$^5$&1.6$\times$10$^{23}$&Y\\
5&345.193&1.026&11&0.3&3.5$\times$10$^2$&3.3$\times$10$^5$&2.3$\times$10$^{23}$&Y\\
6&345.385&1.428&9.8&0.3&3.3$\times$10$^2$&3.3$\times$10$^5$&2.2$\times$10$^{23}$&Y\\
7&345.006&1.820&12&0.4&4.1$\times$10$^2$&2.4$\times$10$^5$&1.9$\times$10$^{23}$&Y\\
8&345.179&1.043&9.1&0.5&3.0$\times$10$^2$&8.2$\times$10$^4$&8.4$\times$10$^{22}$&Y\\
9&345.213&1.022&9.4&0.4&3.1$\times$10$^2$&1.6$\times$10$^5$&1.3$\times$10$^{23}$&Y\\
10&344.992&1.821&10&0.5&3.4$\times$10$^2$&8.0$\times$10$^4$&8.6$\times$10$^{22}$&Y\\
11&345.115&1.592&6.6&0.5&2.2$\times$10$^2$&7.4$\times$10$^4$&7.0$\times$10$^{22}$&Y\\
12&345.716&0.818&7.2&0.5&2.4$\times$10$^2$&7.5$\times$10$^4$&7.3$\times$10$^{22}$&Y\\
13&345.337&1.018&7.2&0.5&2.4$\times$10$^2$&8.3$\times$10$^4$&7.8$\times$10$^{22}$&Y\\
14&344.976&1.783&5.7&0.4&1.9$\times$10$^2$&9.3$\times$10$^4$&7.8$\times$10$^{22}$&N\\
15&344.980&1.752&3.5&0.3&1.2$\times$10$^2$&9.6$\times$10$^4$&6.8$\times$10$^{22}$&N\\
16&345.367&1.434&3.8&0.3&1.2$\times$10$^2$&1.3$\times$10$^5$&8.5$\times$10$^{22}$&Y\\
17&345.296&1.452&10&0.6&3.4$\times$10$^2$&4.4$\times$10$^4$&5.8$\times$10$^{22}$&Y\\
18&345.401&1.419&4.7&0.4&1.6$\times$10$^2$&1.1$\times$10$^5$&8.1$\times$10$^{22}$&N\\
19&345.418&1.401&2.9&0.3&95&1.9$\times$10$^5$&1.0$\times$10$^{23}$&N\\
20&345.395&1.380&4.5&0.5&1.5$\times$10$^2$&4.0$\times$10$^4$&4.1$\times$10$^{22}$&Y\\
21&345.411&1.406&2.9&0.3&96&1.9$\times$10$^5$&1.0$\times$10$^{23}$&Y\\
22&345.434&1.405&2.8&0.3&94&1.1$\times$10$^5$&7.0$\times$10$^{22}$&N\\
23&345.443&1.387&3.5&0.4&1.2$\times$10$^2$&7.9$\times$10$^4$&6.0$\times$10$^{22}$&N\\
24&345.453&1.386&3.2&0.4&1.1$\times$10$^2$&7.4$\times$10$^4$&5.5$\times$10$^{22}$&N\\
25&345.406&1.394&2.9&0.3&95&1.3$\times$10$^5$&7.6$\times$10$^{22}$&Y\\
26&345.389&1.531&2.0&0.3&66&1.2$\times$10$^5$&6.5$\times$10$^{22}$&Y\\
27&345.554&1.508&1.7&0.2&58&1.4$\times$10$^5$&7.0$\times$10$^{22}$&N\\
28&345.398&1.538&1.7&0.2&55&1.4$\times$10$^5$&6.7$\times$10$^{22}$&N\\
29&345.244&1.021&1.8&0.3&58&9.9$\times$10$^4$&5.5$\times$10$^{22}$&N\\
30&345.301&1.040&1.8&0.3&60&1.2$\times$10$^5$&6.3$\times$10$^{22}$&Y\\
31&345.525&1.567&1.8&0.3&59&6.4$\times$10$^4$&4.1$\times$10$^{22}$&N\\
32&345.396&1.389&1.5&--&49&--&--&Y\\
33&345.251&1.038&2.1&0.3&68&1.1$\times$10$^5$&6.3$\times$10$^{22}$&N\\
34&344.878&1.437&3.9&0.5&1.3$\times$10$^2$&4.1$\times$10$^4$&4.0$\times$10$^{22}$&Y\\
35&345.353&1.448&2.5&0.5&82&3.1$\times$10$^4$&2.8$\times$10$^{22}$&Y\\
36&345.236&1.040&2.1&0.4&70&5.3$\times$10$^4$&3.9$\times$10$^{22}$&N\\
37&344.946&1.796&1.6&0.4&54&3.7$\times$10$^4$&2.8$\times$10$^{22}$&N\\
38&345.442&1.558&1.2&0.2&41&1.1$\times$10$^5$&5.3$\times$10$^{22}$&N\\
39&345.257&1.073&1.6&0.3&54&7.4$\times$10$^4$&4.4$\times$10$^{22}$&Y\\
40&345.368&1.036&1.6&0.5&53&1.9$\times$10$^4$&1.8$\times$10$^{22}$&Y\\
41&344.942&1.241&0.55&--&18&--&--&N\\
42&345.539&1.567&1.5&0.3&49&4.5$\times$10$^4$&3.1$\times$10$^{22}$&N\\
43&345.254&1.052&1.9&0.4&64&5.2$\times$10$^4$&3.7$\times$10$^{22}$&N\\
44&344.904&1.803&1.9&0.4&62&4.3$\times$10$^4$&3.3$\times$10$^{22}$&N\\
45&345.502&0.842&0.88&0.3&29&4.4$\times$10$^4$&2.6$\times$10$^{22}$&N\\
46&345.559&1.531&0.85&0.3&28&5.2$\times$10$^4$&2.8$\times$10$^{22}$&N\\
47&345.360&1.389&0.80&0.2&26&6.2$\times$10$^4$&3.1$\times$10$^{22}$&Y\\
48&345.441&0.206&1.8&0.5&61&2.2$\times$10$^4$&2.0$\times$10$^{22}$&Y\\
49&345.137&1.564&0.61&0.2&20&5.7$\times$10$^4$&2.7$\times$10$^{22}$&Y\\
50&345.312&1.046&0.95&0.3&31&5.0$\times$10$^4$&2.9$\times$10$^{22}$&Y\\
51&345.445&1.371&1.3&0.5&42&1.5$\times$10$^4$&1.4$\times$10$^{22}$&N\\
52&345.334&1.428&1.5&0.4&49&3.6$\times$10$^4$&2.6$\times$10$^{22}$&Y\\
53&345.563&1.486&1.1&0.3&35&6.3$\times$10$^4$&3.5$\times$10$^{22}$&N\\
54&344.929&1.836&0.50&--&17&--&--&N\\
55&345.335&1.434&1.1&0.5&36&1.1$\times$10$^4$&1.1$\times$10$^{22}$&Y\\
56&345.434&1.445&1.00&0.3&33&7.5$\times$10$^4$&3.8$\times$10$^{22}$&N\\
57&345.211&1.049&0.55&0.3&18&3.7$\times$10$^4$&2.0$\times$10$^{22}$&N\\
58&345.968&0.598&1.5&0.4&49&3.0$\times$10$^4$&2.3$\times$10$^{22}$&N\\
59&344.945&1.229&0.43&--&14&--&--&N\\
60&345.006&1.532&1.1&0.4&36&2.5$\times$10$^4$&1.9$\times$10$^{22}$&Y\\
61&345.265&1.085&1.2&0.3&39&3.6$\times$10$^4$&2.5$\times$10$^{22}$&N\\
62&344.952&1.195&0.88&0.3&29&6.6$\times$10$^4$&3.3$\times$10$^{22}$&N\\
63&345.337&1.037&0.99&0.4&33&1.7$\times$10$^4$&1.4$\times$10$^{22}$&Y\\
64&345.436&1.418&0.67&0.3&22&4.0$\times$10$^4$&2.2$\times$10$^{22}$&N\\
65&345.501&0.418&0.66&0.2&22&5.4$\times$10$^4$&2.7$\times$10$^{22}$&N\\
66&345.457&0.428&0.81&0.3&27&3.6$\times$10$^4$&2.2$\times$10$^{22}$&N\\
67&344.933&1.251&0.73&0.2&24&7.1$\times$10$^4$&3.3$\times$10$^{22}$&N\\
68&345.064&1.747&0.57&0.2&19&6.0$\times$10$^4$&2.7$\times$10$^{22}$&N\\
69&345.501&0.389&1.3&0.5&42&1.2$\times$10$^4$&1.2$\times$10$^{22}$&N\\
70&345.468&1.431&0.33&--&11&--&--&Y\\
71&345.558&1.522&0.40&--&13&--&--&N\\
72&345.010&1.764&0.60&0.3&20&1.7$\times$10$^4$&1.2$\times$10$^{22}$&N\\
73&345.590&1.491&0.19&--&6.2&--&--&N\\
74&344.932&1.231&0.23&--&7.5&--&--&N\\
75&344.927&1.806&0.82&0.3&27&3.0$\times$10$^4$&1.9$\times$10$^{22}$&N\\
76&344.947&1.813&0.48&0.4&16&1.1$\times$10$^4$&8.3$\times$10$^{21}$&N\\
77&344.955&1.170&0.72&0.2&24&6.2$\times$10$^4$&3.0$\times$10$^{22}$&N\\
78&344.936&1.245&0.30&--&10&--&--&N\\
79&345.384&1.038&0.68&0.3&23&3.5$\times$10$^4$&2.0$\times$10$^{22}$&N\\
80&345.548&1.480&0.34&--&11&--&--&Y\\
81&345.513&0.407&0.82&0.3&27&4.6$\times$10$^4$&2.6$\times$10$^{22}$&N\\
82&345.563&1.526&0.36&0.2&12&3.2$\times$10$^4$&1.5$\times$10$^{22}$&N\\
83&345.450&1.364&0.34&--&11&--&--&N\\
84&345.429&1.455&0.88&0.3&29&3.7$\times$10$^4$&2.3$\times$10$^{22}$&N\\
85&345.010&1.617&0.45&--&15&--&--&N\\
86&345.001&1.615&1.2&0.5&41&1.3$\times$10$^4$&1.3$\times$10$^{22}$&N\\
87&345.319&1.484&0.68&0.4&22&1.8$\times$10$^4$&1.3$\times$10$^{22}$&Y\\
88&345.519&1.639&0.71&0.3&24&4.9$\times$10$^4$&2.5$\times$10$^{22}$&N\\
89&345.133&1.069&1.3&0.6&42&7.0$\times$10$^3$&8.4$\times$10$^{21}$&N\\
90&345.396&1.519&0.42&--&14&--&--&Y\\
91&345.033&1.632&0.76&0.3&25&3.7$\times$10$^4$&2.2$\times$10$^{22}$&N\\
92&345.476&1.568&0.31&--&10&--&--&N\\
93&344.966&1.181&0.41&--&14&--&--&N\\
94&345.240&0.390&0.49&--&16&--&--&N\\
95&345.524&0.404&0.46&0.2&15&3.8$\times$10$^4$&1.9$\times$10$^{22}$&N\\
96&345.590&0.374&0.40&--&13&--&--&Y\\
97&345.031&1.781&0.56&0.4&19&9.6$\times$10$^3$&8.0$\times$10$^{21}$&N\\
98&345.451&0.435&0.39&--&13&--&--&N\\
99&345.078&1.786&0.64&0.3&21&1.9$\times$10$^4$&1.3$\times$10$^{22}$&N\\
100&345.854&1.415&0.47&--&16&--&--&N\\
101&345.217&0.999&0.51&0.3&17&1.8$\times$10$^4$&1.2$\times$10$^{22}$&N\\
102&344.936&1.596&0.44&--&15&--&--&N\\
103&345.065&1.615&1.3&0.5&44&1.7$\times$10$^4$&1.5$\times$10$^{22}$&N\\
104&345.338&1.463&0.73&0.5&24&8.8$\times$10$^3$&8.2$\times$10$^{21}$&N\\
105&345.585&1.484&0.41&0.2&14&3.9$\times$10$^4$&1.8$\times$10$^{22}$&N\\
106&345.854&1.421&0.29&--&9.5&--&--&N\\
107&345.062&1.842&0.53&0.2&18&5.3$\times$10$^4$&2.4$\times$10$^{22}$&N\\
108&345.477&1.563&0.22&--&7.2&--&--&N\\
109&345.454&1.358&0.16&--&5.4&--&--&N\\
110&345.959&0.608&0.33&--&11&--&--&N\\
111&344.949&1.213&0.27&--&8.8&--&--&N\\
112&344.942&1.589&0.45&--&15&--&--&N\\
113&345.539&1.488&0.29&0.3&9.5&1.2$\times$10$^4$&7.4$\times$10$^{21}$&N\\
114&345.576&0.258&0.26&--&8.5&--&--&N\\
115&345.681&0.318&0.43&--&14&--&--&N\\
116&345.330&1.049&0.45&0.2&15&3.9$\times$10$^4$&1.9$\times$10$^{22}$&N\\
117&345.776&1.445&0.34&--&11&--&--&N\\
118&345.511&1.578&0.40&0.3&13&1.6$\times$10$^4$&9.8$\times$10$^{21}$&N\\
119&344.909&1.182&0.65&0.3&22&2.6$\times$10$^4$&1.6$\times$10$^{22}$&N\\
120&345.518&0.835&0.34&--&11&--&--&N\\
121&345.873&0.811&0.30&--&10&--&--&N\\
122&345.489&0.430&0.23&--&7.7&--&--&N\\
123&344.951&1.096&0.37&0.2&12&3.7$\times$10$^4$&1.7$\times$10$^{22}$&N\\
124&344.906&1.196&0.38&0.2&13&4.5$\times$10$^4$&2.0$\times$10$^{22}$&N\\
125&345.092&1.733&0.36&--&12&--&--&N\\
126&345.261&1.108&0.42&0.3&14&2.1$\times$10$^4$&1.2$\times$10$^{22}$&N\\
127&345.467&0.419&0.45&0.2&15&4.8$\times$10$^4$&2.2$\times$10$^{22}$&N\\
128&344.801&1.082&0.27&--&8.9&--&--&N\\
129&345.482&1.233&0.23&--&7.5&--&--&N\\
130&345.464&1.443&0.21&--&7.0&--&--&Y\\
131&344.849&1.312&0.24&--&7.9&--&--&N\\
132&345.484&0.856&0.17&--&5.5&--&--&N\\
133&345.304&0.435&0.12&--&3.9&--&--&N\\
134&344.945&1.600&0.64&0.3&21&2.1$\times$10$^4$&1.4$\times$10$^{22}$&N\\
135&344.950&1.587&0.22&--&7.3&--&--&N\\
136&345.272&1.051&0.36&0.3&12&1.7$\times$10$^4$&9.8$\times$10$^{21}$&N\\
137&345.060&1.727&0.37&--&12&--&--&N\\
138&345.479&0.400&0.26&--&8.7&--&--&N\\
139&344.846&1.293&0.21&--&7.1&--&--&N\\
140&345.909&0.514&0.36&0.2&12&3.4$\times$10$^4$&1.6$\times$10$^{22}$&N\\
141&345.812&1.756&0.26&--&8.5&--&--&N\\
142&344.957&1.188&0.24&--&7.9&--&--&N\\
143&345.320&1.417&0.29&0.4&9.6&7.6$\times$10$^3$&5.5$\times$10$^{21}$&Y\\
144&345.869&1.820&0.21&--&6.8&--&--&N\\
145&345.278&1.112&0.35&0.2&12&2.7$\times$10$^4$&1.4$\times$10$^{22}$&N\\
146&345.634&0.770&0.33&0.2&11&3.5$\times$10$^4$&1.6$\times$10$^{22}$&N\\
147&345.044&1.618&0.32&0.3&11&2.4$\times$10$^4$&1.2$\times$10$^{22}$&N\\
148&344.809&1.075&0.12&--&3.8&--&--&N\\
149&345.683&1.459&0.30&0.3&9.9&2.2$\times$10$^4$&1.1$\times$10$^{22}$&N\\
150&345.566&0.267&0.17&--&5.7&--&--&N\\
151&344.925&1.316&0.33&--&11&--&--&N\\
152&345.699&1.508&0.18&--&5.9&--&--&N\\
153&345.286&0.412&0.14&--&4.6&--&--&N\\
154&345.611&0.365&0.27&--&8.8&--&--&N\\
155&344.916&1.826&0.17&--&5.5&--&--&N\\
156&345.472&0.412&0.22&--&7.2&--&--&N\\
157&345.038&1.848&0.28&0.2&9.2&2.9$\times$10$^4$&1.3$\times$10$^{22}$&N\\
158&345.792&1.757&0.17&--&5.7&--&--&N\\
159&345.254&1.012&0.089&--&3.0&--&--&N\\
160&345.708&1.456&0.24&--&8.0&--&--&N\\
161&345.472&1.231&0.71&0.4&24&1.3$\times$10$^4$&1.1$\times$10$^{22}$&N\\
162&345.497&1.437&0.19&0.3&6.4&8.0$\times$10$^3$&5.0$\times$10$^{21}$&N\\
163&344.783&1.105&0.28&--&9.3&--&--&N\\
164&345.638&0.330&0.11&--&3.7&--&--&N\\
165&345.699&1.853&0.21&--&7.1&--&--&N\\
166&344.986&1.109&0.10&--&3.4&--&--&N\\
167&345.018&1.762&0.17&--&5.7&--&--&Y\\
168&344.893&1.335&0.45&0.4&15&1.0$\times$10$^4$&7.8$\times$10$^{21}$&N\\
169&344.956&1.348&0.47&0.4&16&6.2$\times$10$^3$&5.6$\times$10$^{21}$&N\\
170&345.948&0.723&0.14&--&4.5&--&--&N\\
171&344.891&1.345&0.20&--&6.6&--&--&N\\
172&344.924&1.824&0.19&--&6.4&--&--&N\\
173&344.929&1.610&0.15&--&5.0&--&--&N\\
174&345.247&0.385&0.18&--&5.9&--&--&N\\
175&345.547&0.960&0.13&--&4.4&--&--&N\\
176&344.892&1.847&0.25&--&8.3&--&--&N\\
177&345.888&1.414&0.26&0.2&8.6&2.3$\times$10$^4$&1.1$\times$10$^{22}$&N\\
178&345.510&0.159&0.095&--&3.2&--&--&N\\
179&345.418&1.469&0.15&0.2&4.8&1.7$\times$10$^4$&7.5$\times$10$^{21}$&N\\
180&344.920&1.419&0.17&--&5.5&--&--&N\\
181&345.079&1.847&0.12&--&4.1&--&--&N\\
182&345.236&1.000&0.20&--&6.6&--&--&N\\
183&345.375&1.571&0.14&--&4.7&--&--&N\\
184&345.684&1.516&0.14&--&4.7&--&--&N\\
185&345.532&1.634&0.13&--&4.3&--&--&N\\
186&345.697&1.459&0.19&--&6.1&--&--&N\\
187&345.501&1.415&0.23&0.3&7.5&1.6$\times$10$^4$&8.2$\times$10$^{21}$&N\\
188&345.816&1.765&0.23&--&7.5&--&--&N\\
189&345.086&1.611&0.32&0.4&11&5.2$\times$10$^3$&4.4$\times$10$^{21}$&N\\
190&345.043&1.838&0.21&0.3&7.0&1.4$\times$10$^4$&7.5$\times$10$^{21}$&N\\
191&345.442&1.572&0.11&--&3.8&--&--&N\\
192&345.468&1.225&0.13&--&4.2&--&--&N\\
193&344.951&1.082&0.34&0.3&11&9.6$\times$10$^3$&6.7$\times$10$^{21}$&N\\
194&345.482&0.431&0.27&0.3&8.8&1.9$\times$10$^4$&9.8$\times$10$^{21}$&N\\
195&344.913&1.165&0.13&--&4.4&--&--&N\\
196&344.933&1.820&0.17&--&5.7&--&--&N\\
197&345.458&1.559&0.13&0.2&4.5&1.1$\times$10$^4$&5.3$\times$10$^{21}$&N\\
198&345.792&1.440&0.096&--&3.2&--&--&N\\
199&345.098&1.726&0.11&--&3.7&--&--&N\\
200&345.293&1.093&0.14&--&4.5&--&--&N\\
201&344.991&1.150&0.11&--&3.8&--&--&N\\
\end{longtable}
\end{center}
}

\end{document}